\documentclass[
 reprint,
 superscriptaddress,
 amsmath,amssymb,
 aps,
 pra,
]{revtex4-2}

\usepackage{graphicx}
\usepackage{dcolumn}
\usepackage{bm}
\usepackage{hyperref}
\usepackage{siunitx}
\usepackage{xspace}
\usepackage{xcolor}

\newcommand{\ket}[1]{\ensuremath{\lvert #1 \rangle}\xspace}%
\newcommand{\avg}[1]{\ensuremath{\langle #1 \rangle}\xspace}%

\begin{document}

\newcommand{\partitle}[1]{\subsection*{#1}}
\newcommand{\figscale}{1.114}

\title{Quantum gas microscopy of Kardar-Parisi-Zhang superdiffusion}

\author{David Wei}
    \affiliation{Max-Planck-Institut f\"{u}r Quantenoptik, 85748 Garching, Germany}
    \affiliation{Munich Center for Quantum Science and Technology (MCQST), 80799 Munich, Germany}

\author{Antonio Rubio-Abadal}
    \altaffiliation[Present address: ]{ICFO – Institut de Ciencies Fotoniques, The Barcelona Institute of Science and Technology, 08860 Castelldefels (Barcelona), Spain}
    \affiliation{Max-Planck-Institut f\"{u}r Quantenoptik, 85748 Garching, Germany}
    \affiliation{Munich Center for Quantum Science and Technology (MCQST), 80799 Munich, Germany}

\author{Bingtian Ye}
    \affiliation{Department of Physics, University of California, Berkeley, California 94720, USA}

\author{Francisco Machado}
    \affiliation{Department of Physics, University of California, Berkeley, California 94720, USA}
    \affiliation{Materials Science Division, Lawrence Berkeley National Laboratory, Berkeley, California 94720, USA}

\author{Jack Kemp}
    \affiliation{Department of Physics, University of California, Berkeley, California 94720, USA}

\author{Kritsana Srakaew}
    \affiliation{Max-Planck-Institut f\"{u}r Quantenoptik, 85748 Garching, Germany}
    \affiliation{Munich Center for Quantum Science and Technology (MCQST), 80799 Munich, Germany}

\author{Simon Hollerith}
    \affiliation{Max-Planck-Institut f\"{u}r Quantenoptik, 85748 Garching, Germany}
    \affiliation{Munich Center for Quantum Science and Technology (MCQST), 80799 Munich, Germany}

\author{Jun Rui}
    \altaffiliation[Present address: ]{Hefei National Laboratory for Physical Sciences at the Microscale, University of Science and Technology of China, Hefei, Anhui 230026, China}
    \affiliation{Max-Planck-Institut f\"{u}r Quantenoptik, 85748 Garching, Germany}
    \affiliation{Munich Center for Quantum Science and Technology (MCQST), 80799 Munich, Germany}

\author{Sarang Gopalakrishnan}
    \affiliation{Department of Physics, The Pennsylvania State University, University Park, Pennsylvania 16802, USA}
    \affiliation{Department of Physics and Astronomy, College of Staten Island, Staten Island, New York 10314, USA}
 
\author{Norman Y. Yao}
    \affiliation{Department of Physics, University of California, Berkeley, California 94720, USA}
    \affiliation{Materials Science Division, Lawrence Berkeley National Laboratory, Berkeley, California 94720, USA}

\author{Immanuel Bloch}
    \affiliation{Max-Planck-Institut f\"{u}r Quantenoptik, 85748 Garching, Germany}
    \affiliation{Munich Center for Quantum Science and Technology (MCQST), 80799 Munich, Germany}
    \affiliation{Fakult\"{a}t f\"{u}r Physik, Ludwig-Maximilians-Universit\"{a}t, 80799 Munich, Germany}

\author{Johannes Zeiher}
    \affiliation{Max-Planck-Institut f\"{u}r Quantenoptik, 85748 Garching, Germany}
    \affiliation{Munich Center for Quantum Science and Technology (MCQST), 80799 Munich, Germany}

\date{\today}

\begin{abstract}
    
    
    The Kardar-Parisi-Zhang (KPZ) universality class describes the coarse-grained behavior of a wealth of classical stochastic models. Surprisingly, it was recently conjectured to also describe spin transport in the one-dimensional quantum Heisenberg model.
    We test this conjecture by experimentally probing transport in a cold-atom quantum simulator via the relaxation of domain walls in spin chains of up to 50 spins. We find that domain-wall relaxation is indeed governed by the KPZ dynamical exponent $z = 3/2$, and that the occurrence of KPZ scaling requires both integrability and a non-abelian SU(2) symmetry.
    Finally, we leverage the single-spin-sensitive detection enabled by the quantum-gas microscope to measure a novel observable based on spin-transport statistics, which yields a clear signature of the non-linearity that is a hallmark of KPZ universality.
    
\end{abstract}

\maketitle


Hydrodynamics captures the evolution of a system from local to global equilibrium~\cite{Spohn2012,Birkhoff2015}.
In many-particle systems, the conventional lore is that---upon coarse-graining---such hydrodynamics naturally emerges from the microscopic equations of motion of both classical and quantum systems~\cite{Wyatt2006,Mukerjee2006,Erdos2007,Castro-Alvaredo2016,Zu2021}.
A celebrated example of the emergence of hydrodynamics is the so-called Kardar-Parisi-Zhang (KPZ) equation, which governs a wealth of disparate phenomena ranging from interface growth to the shapes of polymers and the propagation of shock waves~\cite{Kardar1986,Halpin2015,Spohn2020}.
A single equation can describe so many distinct physical systems because of the notion of \emph{universality}~\cite{Kadanoff1990}.
The canonical examples of KPZ hydrodynamics are classical systems subject to disorder and noise, or alternatively, interacting Galilean fluids~\cite{Gubinelli2017}.

A tremendous amount of recent excitement has centered upon the prediction that KPZ hydrodynamics should emerge in an entirely distinct setting---a one-dimensional spin-1/2 quantum Heisenberg chain~\cite{Znidaric2011a,Ljubotina2017,Ljubotina2019,Gopalakrishnan2019a,DeNardis2019,Gopalakrishnan2019,Bulchandani2020a}.
The appearance of KPZ hydrodynamics in this context is particularly surprising, since quantum magnets are neither subject to extrinsic noise nor Galilean invariant.
Indeed, conventional hydrodynamics would predict that spin transport in such a system is diffusive; however, the Heisenberg model is integrable, and has stable, ballistically propagating quasiparticles~\cite{Castro-Alvaredo2016,Bertini2016,Ilievski2017,Bulchandani2018}.
The subtle interplay between this integrability and the model's SU(2) symmetry leads to anomalous \emph{superdiffusive} spin transport~\cite{Prosen2013,Das2019,Krajnik2020,Bulchandani2020a,Krajnik2020a,Ilievski2020}.
To date, a full theory of KPZ hydrodynamics in the Heisenberg model remains elusive~\cite{Bertini2021,Bulchandani2021} and experimentally characterizing the nature of the anomalous spin transport is a subject of widespread effort~\cite{Hild2014,Jepsen2020,Jepsen2021,Scheie2021}.


\begin{figure*}[t!]
    \centering
    \includegraphics[scale=\figscale]{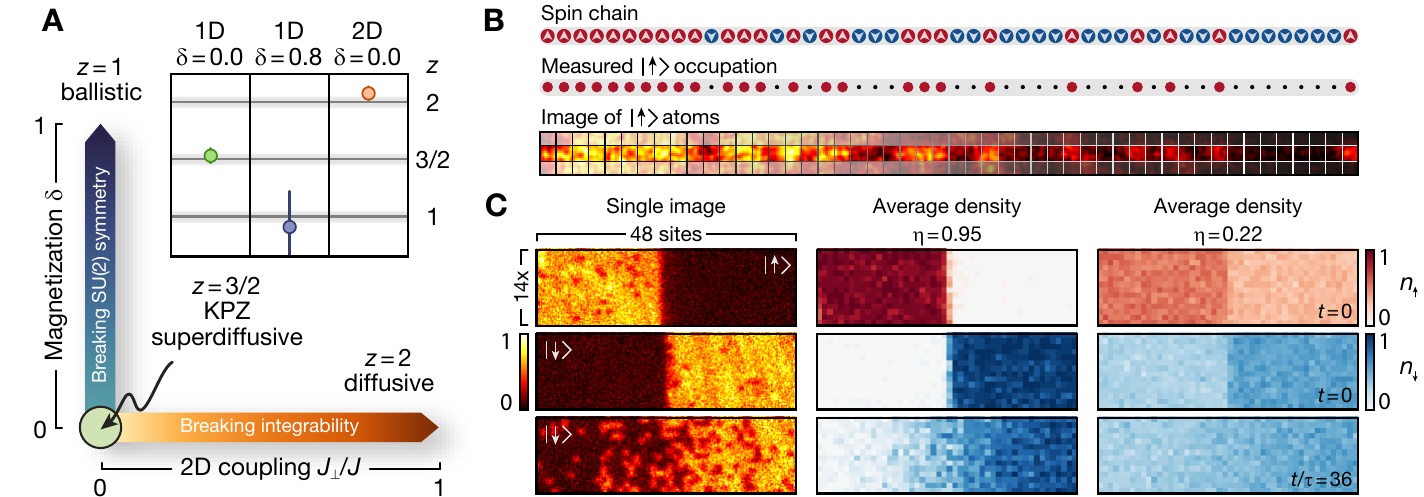}
    \caption{\textbf{Hydrodynamic transport in Heisenberg chains and schematic of the experimental system.}
        (\textbf{A}) Dynamical exponents for finite-temperature Heisenberg chains. Whereas integrable systems typically display ballistic transport (magnetized chains, $\delta > 0$), non-integrable systems are generically diffusive (2D Heisenberg model, $J_\perp > 0$). For unmagnetized Heisenberg chains, transport is expected to fall into the KPZ universality class with a superdiffusive exponent $z = 3/2$.
        (Inset) By measuring polarization transfer $P (t)$ across a domain wall, we directly observe these transport regimes: superdiffusion in the unmagnetized case (green), ballistic transport at finite net magnetization (blue), and diffusion in 2D (orange). Exponents are extracted by fitting $P (t) \propto t^{1/z}$; for the ballistic case we additionally fit a vertical intercept to account for transient initial-time dynamics. Error bars denote the standard deviation (s.d.) of the fit.
        (\textbf{B}) In each experimental run, we measure the spin states of a Heisenberg chain (top) by removing one spin species (center) and imaging the atomic site occupation (bottom).
        (\textbf{C}) The Heisenberg chains are realized in a 2D atomic Mott insulator (analysis region depicted) with controllable inter-chain coupling. Our setup allows us to prepare domain walls with high purity $\eta$ (left, center column) and low purity $\eta$ (right). We measure the time evolution of both $\ket{\uparrow}$ (top) and $\ket{\downarrow}$ (center, bottom row) atoms to extract the polarization transfer.
    }
    \label{fig:1}
\end{figure*}

In this work, we explore the superdiffusive dynamics of the ferromagnetic Heisenberg model using a quantum-gas microscope with single-site resolution and single-spin-sensitive detection in spin chains of up to $50$ spins.
Our main results are threefold.
First, we observe superdiffusive spin transport with the dynamical exponent $z=3/2$, consistent with KPZ hydrodynamics.
Second, we demonstrate that both integrability and a non-abelian symmetry are essential for observing superdiffusion: Breaking integrability by tuning dimensionality restores diffusion, while breaking the symmetry by preparing an initial state with net magnetization leads to ballistic transport (Fig.~\ref{fig:1}A).
Finally, leveraging the ability of our experimental setup to detect spin-resolved \emph{snapshots} of the entire sample, we map the shot-to-shot dynamical fluctuations (i.e., the ``full counting statistics'') of the magnetization.
These fluctuations carry clear signatures of the intrinsic non-linearity associated with KPZ hydrodynamics~\cite{Hartmann2018}, and distinguish it from other potential mechanisms for superdiffusion such as L\'evy flights~\cite{Bulchandani2021}.

\partitle{Experimental system}


In our experiment, we probed the transport dynamics of bosonic $^{87}$Rb atoms trapped in an optical lattice; the atoms occupy the two hyperfine ground states $\ket{\uparrow} = \ket{F=1, m_F=-1}$ and $\ket{\downarrow} = \ket{F=2, m_F=-2}$ and their dynamics are captured by a two-species Bose-Hubbard model with on-site interaction $U$ and tunnel coupling $\tilde{t}$.
At unit filling and in the limit of strong interactions, the direct tunneling between lattice sites is suppressed and spin dynamics occur via second-order spin-exchange.
The system can be mapped to the spin-1/2 XXZ model for $\ket{\uparrow}$ and $\ket{\downarrow}$~\cite{Duan2003,Kuklov2003}, and, in one dimension (1D), is described by the Hamiltonian
\begin{align}
    \hat{H} = -J \sum_{j}\left( \hat{S}^x_j\hat{S}^x_{j+1}+\hat{S}^y_j\hat{S}^y_{j+1} + \Delta \hat{S}^z_j \hat{S}^z_{j+1}\right),
    \label{eq:XXZ}
\end{align}
where $\Delta$ quantifies the interaction anisotropy and $J = 4 \, \tilde{t}^2 / U$ characterizes the spin-exchange coupling.
In our system, the atomic scattering properties yield $\Delta \approx 1$ and the system maps to the isotropic ferromagnetic Heisenberg model~\cite{SI}.

We began our experiment by loading a spin-polarized 2D degenerate gas of approximately $2000$ atoms into a square optical lattice with a spacing of $a = \SI{532}{\nano\metre}$.
We realized a homogeneous box potential over $50\,\times\,22$ sites by additionally projecting light at a wavelength of $\SI{670}{\nano\metre}$ with a digital micromirror device (DMD), preparing a Mott insulator with a filling of $n_0 = 0.93(1)$ in this box (see details in~\cite{SI}).
Local spin control was realized using light at a wavelength of $\SI{787}{\nano\metre}$ on the DMD~\cite{Fukuhara2013} to apply a site-resolved differential light shift between $\ket{\uparrow}$ and $\ket{\downarrow}$; subsequent microwave driving allows for local flips of the spatially addressed spins.

Such quantum control enabled us to prepare spin domain walls~\cite{Halimeh2014,Ljubotina2017,Misguich2017,Ljubotina2019} by spatially addressing half the system.
Subsequently, we prepared \emph{high-entropy} states by globally rotating the spins away from the $S^z$-axis using a resonant microwave pulse and then locally dephasing them by projecting a site-to-site random spin-dependent potential, which we modified from shot to shot~\cite{SI} (Fig.~\ref{fig:1}C).
More precisely, our experiments focused on tracking spin dynamics starting from a class of initial states comprising a spin domain wall with magnetization difference $2\eta$ in the middle of the spin chain: i.e., one half of the system has magnetization $\eta$ and the other half of the system has magnetization $-\eta$.
In the infinite-temperature limit, $\eta \to 0$, the relaxation of such states yields linear response transport coefficients, as the derivative of the spin profile is precisely the dynamical spin structure factor~\cite{Ljubotina2017,Ljubotina2019}.

In order to probe 1D spin dynamics in our system, we rapidly quenched the lattice depth along 1D tubes comprising $50$ sites, which suddenly increased the spin-exchange coupling from zero to $J/\hbar = \SI{64(1)}{s^{-1}}$.
After tracking the spin dynamics for up to $\sim 45$ spin-exchange times $\tau= \hbar/J$, we removed one spin component and measured the remaining occupation via fluorescence imaging (Fig.~\ref{fig:1}B).

\partitle{Superdiffusive spin transport}


\begin{figure*}
    \centering
    \includegraphics[scale=\figscale]{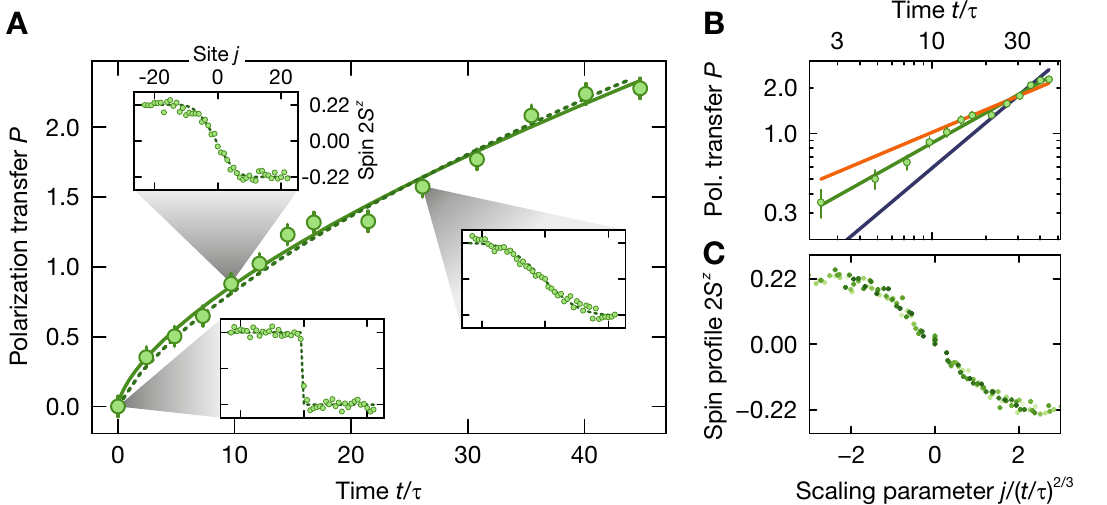}
    \caption{\textbf{Superdiffusive spin transport in a high-temperature Heisenberg chain.}
        (\textbf{A}) The polarization transfer for a domain-wall initial state with a contrast of $\eta = 0.22$ grows as a power law ($P(t) \propto t^{1/z}$) with a fitted exponent $z = 1.54 (7)$ (solid line), indicating superdiffusive transport. The experimental data agrees well with numerical Heisenberg-model simulations (dashed line).
        The insets show the averaged spin profiles $2 S_j^z (t)$ at times $ t / \tau = 0, 10, 26$, which are compared to simulations (dashed lines).
        (\textbf{B}) Polarization transfer in a double-logarithmic plot. The solid lines are power-law fits with fixed exponents, where a distinction between $z = 3/2$ (green) and both $z = 2$ (brown) and $z = 1$ (blue) is visible.
        (\textbf{C}) When rescaling time by the inverse dynamical exponent, the spatial spin profiles at times $t/\tau = 5$ to $35$ (light to dark green) collapse to a characteristic shape consistent with the integrated KPZ function.
        Error bars denote the standard error of the mean (s.e.m).
    }
    \label{fig:2}
\end{figure*}

To explore the nature of anomalous spin transport in the 1D Heisenberg model, we initialize the spins in a high-entropy domain-wall state with $\eta=0.22$.
We characterize the subsequent spin transport by measuring the polarization transfer, $P(t)$, defined as the total number of spins which have crossed the domain wall by time $t$~\cite{SI}.
The emergence of hydrodynamics is characterized by the power-law scaling of $P(t)\sim t^{1/z}$, and immediately enables us to extract the underlying dynamical exponent $z$.
As depicted in Fig.~\ref{fig:2}A, the data exhibit a superdiffusive exponent, $z = 1.54(7)$, consistent with KPZ scaling.
By comparison, neither a diffusive ($z=2$) nor ballistic ($z=1$) exponent accurately capture the observed dynamics (Fig.~\ref{fig:2}B).

To further explore the superdiffusive dynamics, we investigate the spatially resolved spin profiles.
Our experimental observations are in quantitative agreement with simulations based upon novel tensor-network numerical techniques~\cite{SI,White2018,Ye2020} and conform to KPZ dynamics (Fig.~\ref{fig:2}A).
Crucially, when appropriately rescaled by the dynamical exponent, all of the observed spatio-temporal profiles collapse onto a scaling form consistent with the KPZ scaling function (Fig.~\ref{fig:2}C).

Somewhat surprisingly, we also observe a superdiffusive exponent of $z = 1.45(4)$~\cite{SI} upon changing the initial state to a near-pure domain wall with $\eta = 0.95$.
While this contrasts with the analytical expectations for a pure initial state ($\eta = 1$), which has been shown to exhibit logarithmically corrected diffusion at asymptotically late timescales~\cite{Misguich2017,Gamayun2019,Ye2020,Bulchandani2021}, it is consistent with finite-time numerics, which find a robust superdiffusive exponent for a large range of $\eta$ values~\cite{SI}.

\partitle{Breaking integrability and SU(2) symmetry}

To probe the microscopic origin of the emergent superdiffusive transport, it is instructive to consider the transport dynamics on top of a small net magnetization background~\cite{Gopalakrishnan2019,Gopalakrishnan2019a,DeNardis2019,DeNardis2020}.
In our experiments, this corresponds to preparing domain walls with a finite overall magnetization $\delta$, i.e. one half of the system has a magnetization $\eta + \delta$ and the other half $-\eta + \delta$.
Stable quasiparticles then render spin transport ballistic (Fig.~\ref{fig:1}A), leading to a characteristic polarization-transfer rate which scales linearly with net magnetization $\delta$~\cite{Gopalakrishnan2019}.
Even when $\delta = 0$ on average, random local fluctuations of the magnetization will be present; thus, the net magnetization in a typical region of size $\ell$ will scale as $1/\sqrt{\ell}$.
Therefore, the average spin transport rate across a region of size $\ell$ also scales as $1/\sqrt{\ell}$, implying that the transport time across the region scales as $\ell / (1/\sqrt{\ell}) \sim \ell^{3/2}$, precisely yielding the KPZ exponent $z=3/2$ (see details in~\cite{SI}).

This intuitive analysis suggests two key requirements for superdiffusive transport:
(i) integrability ensures the presence of stable quasiparticles that move ballistically, and
(ii) the presence of a non-abelian SU(2) symmetry makes the characteristic velocity of the ballistic contribution to spin transport vanish.
By tuning the dimensionality or the net magnetization of the initial state, we remove each of these ingredients individually and study the resulting spin dynamics.

\begin{figure}
    \centering
    \includegraphics[scale=\figscale]{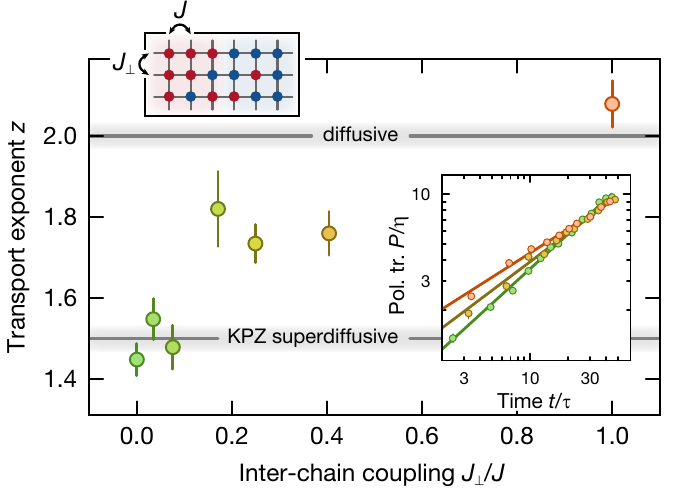}
    \caption{\textbf{Evolution towards diffusive transport under a breakdown of integrability.}
        Fitted power-law exponent $z$ for the spin polarization transfer at different coupling strengths between individual 1D chains with initial domain walls with $\eta \sim 0.9$. Starting from superdiffusive transport in the purely 1D case, $z = 1.45 (5)$, increased inter-chain coupling breaks the integrability of the system and leads to a crossover towards diffusive transport, reaching $z = 2.08 (4)$ in the 2D case, as generically expected for non-integrable systems.
        The inset depicts the normalized polarization transfer $P(t) / \eta$ for $J_\perp / J = 0, 0.4$ and $1$ (green to orange).
        Error bars denote s.d. of the fit.
    }
    \label{fig:3}
\end{figure}

To break integrability, we turn on a finite inter-chain coupling $J_\perp$ by lowering the lattice depth orthogonal to the 1D spin chains, which effectively causes the system to become 2D~\cite{Tang2018,Nichols2019}.
We measure the dependence of the polarization transfer on the inter-chain coupling, starting from an unmagnetized domain wall ($\eta \approx 0.9$, $\delta = 0$). As shown in Fig.~\ref{fig:3}, the extracted dynamical exponents exhibit a clear flow from superdiffusive transport when $J_\perp = 0$ to diffusive transport, $z = 2.08(4)$, when $J_\perp = J$.
Interestingly, for $J_\perp / J \lesssim 0.1$ integrability is strictly broken but the transport dynamics remain consistent with superdiffusion within experimentally accessible timescales.
This observation bolsters recent theoretical expectations, which suggest that superdiffusion can be particularly robust to perturbations that do not break the non-abelian symmetry~\cite{DeNardis2021}.

\begin{figure}
    \centering
    \includegraphics[scale=1.2]{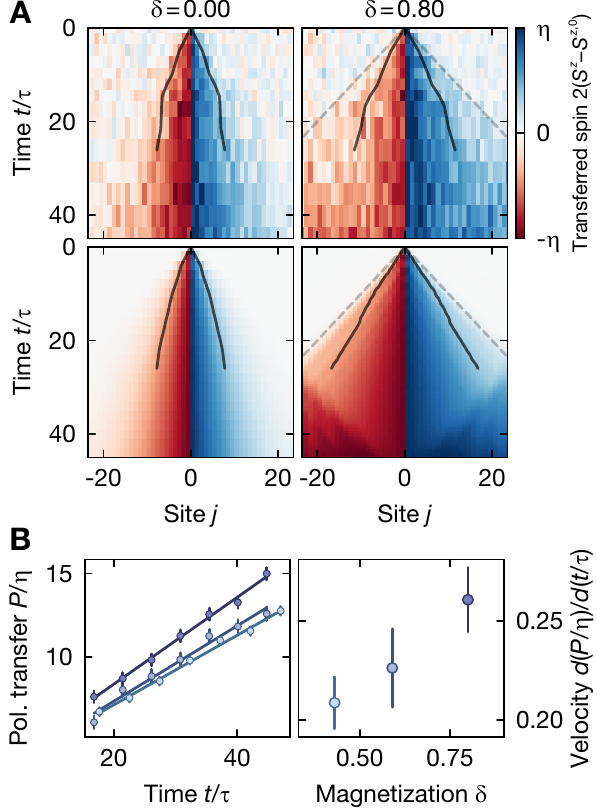}
    \caption{\textbf{Ballistic spin dynamics under broken SU(2) symmetry.}
        (\textbf{A}) Averaged experimental (top) and numerical (bottom) spin profiles $S_j^z (t)$, from which the initial profile $S_j^{z,0}$ is subtracted.
        (Left) Unmagnetized low-purity domain wall, $\delta = 0, \eta = 0.22$ (from Fig.~\ref{fig:2}). Spin transport results from the increase of the spin profile width, which scales with the superdiffusive dynamical exponent. The numerical simulation is performed for an ideal Heisenberg chain at $\delta = 0, \eta = 0.1$. The black lines indicate the position where the spin profile crosses $2 S^z(t) = 0.4\,\eta$ and follows the $z = 3/2$ scaling.
        (Right) Magnetized domain wall, $\delta = 0.80, \eta = 0.12$. At the outer edge the contribution of magnons is visible, transporting spin with the independently measured speed of the light cone (dashed line). The majority of the spin is carried by quasiparticles within the light cone, leading to the width of the profile growing faster than in the unmagnetized case (solid line).
        The numerical simulation at $\delta = 0.8, \eta = 0.2$ shows a qualitatively similar behavior. At $t/\tau = 25$ the magnons reach the system edge and are reflected.
        (\textbf{B}) To extract the ballistic polarization-transfer velocity, we linearly fit the normalized polarization transfer after a crossover time, $t/\tau > 16$ (left). We observe a growth of the transfer velocity when increasing the initial domain-wall magnetization $\delta$ (right, light to dark blue).
        Error bars denote s.d. of the fit.
    }
    \label{fig:4}
\end{figure}

Next, let us explore the effect of breaking the underlying SU(2) symmetry using initial states with finite net magnetization $\delta$~\cite{SI}.
Working with an imbalanced domain-wall initial state ($\eta=0.12$, $\delta=0.80$), we observe two main differences compared to the unmagnetized ($\delta = 0$) case (Fig.~\ref{fig:4}A).
First, the polarization profile exhibits a fast ballistic component that follows the light cone of the dynamics ($j = t/\tau$, dashed line in Fig.~\ref{fig:4}A).
This contribution arises from the fastest quasiparticles which now transport spin above the magnetized background~\cite{Weiner2020}.
Second, within this light cone, polarization also spreads substantially faster compared to the unmagnetized case;
this comprises the bulk of the spin transport and is mediated by slower-moving, net-magnetization-carrying quasiparticles.

At early times, the polarization-transfer dynamics exhibit a superdiffusive power law, before crossing over to linear ballistic transport at later times~\cite{SI}.
In particular, by fitting a power law to the late-time data, $t/\tau > 16$, we extract a dynamical exponent $z = 0.9(3)$, consistent with ballistic spin transport (Fig.~\ref{fig:4}B).
While our results agree qualitatively with numerical simulations of the Heisenberg model, the magnitude of the measured polarization transfer is smaller; this can be understood as resulting from the presence of hole defects in the initial state~\cite{SI,Fava2020}.
In addition to verifying the ballistic nature of the spin dynamics, we can also directly extract the velocity of the underlying quasiparticles;
by controlling the overall magnetization of the initial state, we observe the expected increase of the velocity with $\delta$ (Fig.~\ref{fig:4}B), an essential cornerstone for understanding the presence of KPZ superdiffusion in spin chains~\cite{Gopalakrishnan2019}.


\begin{figure*}
    \centering
    \includegraphics[scale=\figscale]{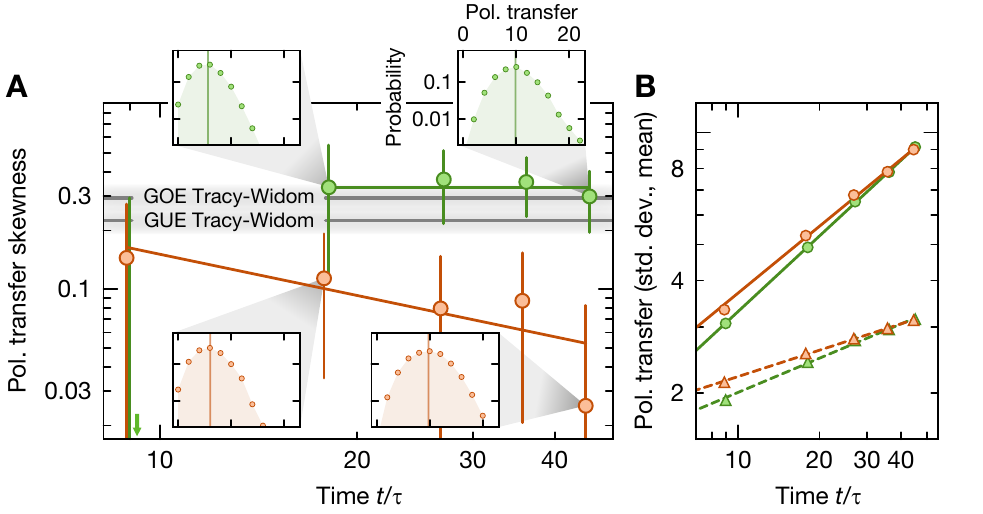}
    \caption{\textbf{Distribution function of polarization transfer.}
        (\textbf{A}) The probability distribution asymmetry of the polarization transfer expected for KPZ transport is quantified by the skewness. We compare the pure domain-wall dynamics in the 1D case (green) with the non-integrable 2D case at $J_\perp / J = 0.25$ (orange). Whereas the 2D case becomes symmetric at late times, the 1D distribution remains asymmetric with a skewness of $0.33(8)$. Gray lines indicate the skewness of the GOE and Gaussian-unitary-ensemble (GUE) TW distributions~\cite{SI,Prahofer2000}. Colored lines serve as guides to the eye.
        (Insets) Probability distributions of the polarization transfer in a logarithmic scale. The vertical line marks the mean of the distribution.
        (\textbf{B}) The mean (circles) of the polarization transfer is consistent with the data shown in Fig.~\ref{fig:3} and scales with the power-law (solid lines) exponent $1/z = 0.67(1)$ in 1D; $1/z = 0.60(2)$ in 2D.
        The standard deviation (triangles) features another characteristic transport exponent (the growth exponent~\cite{Family1985}) which agrees with the extracted power-law (dashed lines) exponent, $\beta = 0.31(1)$ in 1D; $\beta = 0.24(1)$ in 2D. 
        Error bars denote the s.d. obtained from a bootstrap analysis.
    }
    \label{fig:5}
\end{figure*}

\partitle{Observing KPZ hydrodynamics}

Our previous observations have focused on characterizing superdiffusive spin transport;
however, from the perspective of observing KPZ universality, this is insufficient, as multiple different classes of hydrodynamics can exhibit the same dynamical exponent of $z = 3/2$.
To distinguish these classes, we go beyond measurements of the average polarization transfer and analyze the full \emph{distribution function} of the polarization transfer across snapshots.
This distribution function can distinguish KPZ from potential alternatives such as L\'evy flights:
for \emph{all} linear processes (such as L\'evy flights or time-rescaled diffusion) the fluctuations of $P(t)$ at late times are necessarily symmetric about the mean;
for KPZ, the limiting distribution of $P(t)$ is the Tracy-Widom distribution~\cite{SI}, which is strongly asymmetric~\cite{Prahofer2000,Spohn2020}.

Measuring the statistics of the polarization-transfer distribution therefore gives us a direct experimental observable to discern the underlying hydrodynamical transport equations;
this analysis fundamentally relies on the single-shot nature and the single-spin sensitivity of our quantum-gas microscope.
As we measure the occupation of a single spin species per snapshot, we approximate the polarization-transfer statistics by the statistics for the single-species atom-number transfer, $N_T^{\uparrow (\downarrow)} \approx P / 2$, where $N_T^{\uparrow}$ is the number of $\ket{\uparrow}$ atoms on the side of the domain wall initialized with the opposite spin $\ket{\downarrow}$.
We quantify the asymmetry of the distribution by its skewness $(\mu_3(t) - \mu_3(0)) / (\mu_2(t) - \mu_2(0))^{3/2}$~\cite{SI}, where $\mu_k$ denotes the $k$-th central moment of the distribution.

To begin, we characterize the skewness of the polarization transfer starting from a high-purity domain wall ($\eta = 0.89$, $\delta=0$) for a 2D geometry with an inter-chain coupling strength $J_\perp / J = 0.25$.
As a function of time, the skewness of the polarization transfer distribution decays toward zero (Fig.~\ref{fig:5}), exhibiting a clear trend to a fully \emph{symmetric} distribution, consistent with \emph{linear} diffusive processes expected for the non-integrable 2D Heisenberg model.

If the 1D Heisenberg model is actually governed by \emph{non-linear} KPZ hydrodynamics, one expects a markedly distinct behavior for the skewness as a function of time.
In particular, the non-linearity of the KPZ equation would lead to a finite skewness, which is constant over time.
We indeed observe that the skewness saturates to a finite value of $0.33(8)$ when starting from an initial state with $\eta = 0.91$ and $\delta=0$ (Fig.~\ref{fig:5}).
In agreement with numerical simulations, this value is consistent with the skewness of the Gaussian-orthogonal-ensemble (GOE) Tracy-Widom (TW) distribution, $0.294$~\cite{Prahofer2000}, and contrasts with the generic $t^{-1/3}$ power-law decay of the skewness expected for all linear transport equations with $z=3/2$~\cite{SI}.
Directly ruling out linear transport processes, our experiment thus provides a strong indication that transport in the 1D quantum Heisenberg chain is indeed governed by KPZ hydrodynamics.

\partitle{Discussion and Conclusion}


Our results support the theoretical conjecture that spin transport in the 1D Heisenberg model belongs to the KPZ universality class, with a superdiffusive transport exponent $z = 3/2$.
We have experimentally demonstrated that both integrability and a non-abelian symmetry are essential for stabilizing superdiffusive transport.
Moreover, we exploit the single-spin sensitivity of our setup to extract the full distribution function of the polarization transfer.
This distribution function exhibits a large skewness that does not decay in time, demonstrating for the first time, that spin transport in this system belongs to a strongly coupled, non-linear dynamical universality class.

Our work builds and expands upon recent experimental explorations of Heisenberg-model spin dynamics.
These experiments include neutron scattering studies of the quantum material $\textrm{KCuF}_3$~\cite{Scheie2021}, as well as experiments probing the relaxation of spin-spiral initial states in ultracold gases~\cite{Hild2014,Jepsen2020,Jepsen2021}.
In the 1D Heisenberg model, the relaxation of such spin-spiral states is non-generic because they are approximate eigenstates in the long-wavelength limit~\cite{Bulchandani2021}.
Empirically, spin spirals relax with a diffusive exponent $z = 2$.
By considering a more generic family of domain-wall initial states, we are able to directly probe (and controllably move away from) the high-temperature linear-response limit where KPZ transport is conjectured to occur.

Our results open the door to a number of intriguing directions.
First, the discrepancy between the relaxation of domain walls and spin spirals (away from linear response) indicates that relaxation in integrable systems is generally strongly state dependent;
we lack a theory of this non-linear regime.
Second, the robustness of our results along the crossover from the Heisenberg to the (non-integrable) Bose-Hubbard regime remains to be fully understood~\cite{DeNardis2021}.
In this context, a comparison between the non-integrable Bose-Hubbard model and the integrable Fermi-Hubbard model~\cite{Fava2020} could be of particular interest.
Finally, the observable we introduced to capture fluctuation effects---namely, the statistics of single shots of the polarization transfer---promises to be a powerful diagnostic tool for new phases of interacting quantum systems.
Fortuitously, a theory of this quantity already exists for the KPZ universality class;
developing a more general theory of such transport fluctuations is an important task for future theoretical work.

\textit{Note added:} During the completion of this manuscript, we became aware of related work observing superdiffusive transport in a long-range interacting ion chain~\cite{IonPaper}.

\begin{acknowledgments}
    

    We gratefully acknowledge discussions with Joel Moore, Romain Vasseur and Michael Zaletel.
    We thank Toma{\v z} Prosen and Herbert Spohn for comments on the manuscript.
    We acknowledge funding by the Max Planck Society (MPG), the European Union (PASQuanS grant number 817482) and the Deutsche Forschungsgemeinschaft (DFG, German Research Foundation) under Germany’s Excellence Strategy – EXC-2111 – 390814868.
    BY, FM, JK and NYY acknowledge support from the ARO grant no.~W911NF-21-1-0262 and through the MURI program (W911NF-20-1-0136).
    JR acknowledges funding from the Max Planck Harvard Research Center for Quantum Optics.
    SG acknowledges support from the NSF DMR-1653271.
    NYY acknowledges support from the David and Lucile Packard foundation and the W.~M.~Keck foundation.

\end{acknowledgments}


\bibliography{SuperDiffusion}

\begin{thebibliography}{58}%
\makeatletter
\providecommand \@ifxundefined [1]{%
 \@ifx{#1\undefined}
}%
\providecommand \@ifnum [1]{%
 \ifnum #1\expandafter \@firstoftwo
 \else \expandafter \@secondoftwo
 \fi
}%
\providecommand \@ifx [1]{%
 \ifx #1\expandafter \@firstoftwo
 \else \expandafter \@secondoftwo
 \fi
}%
\providecommand \natexlab [1]{#1}%
\providecommand \enquote  [1]{``#1''}%
\providecommand \bibnamefont  [1]{#1}%
\providecommand \bibfnamefont [1]{#1}%
\providecommand \citenamefont [1]{#1}%
\providecommand \href@noop [0]{\@secondoftwo}%
\providecommand \href [0]{\begingroup \@sanitize@url \@href}%
\providecommand \@href[1]{\@@startlink{#1}\@@href}%
\providecommand \@@href[1]{\endgroup#1\@@endlink}%
\providecommand \@sanitize@url [0]{\catcode `\\12\catcode `\$12\catcode
  `\&12\catcode `\#12\catcode `\^12\catcode `\_12\catcode `\%12\relax}%
\providecommand \@@startlink[1]{}%
\providecommand \@@endlink[0]{}%
\providecommand \url  [0]{\begingroup\@sanitize@url \@url }%
\providecommand \@url [1]{\endgroup\@href {#1}{\urlprefix }}%
\providecommand \urlprefix  [0]{URL }%
\providecommand \Eprint [0]{\href }%
\providecommand \doibase [0]{https://doi.org/}%
\providecommand \selectlanguage [0]{\@gobble}%
\providecommand \bibinfo  [0]{\@secondoftwo}%
\providecommand \bibfield  [0]{\@secondoftwo}%
\providecommand \translation [1]{[#1]}%
\providecommand \BibitemOpen [0]{}%
\providecommand \bibitemStop [0]{}%
\providecommand \bibitemNoStop [0]{.\EOS\space}%
\providecommand \EOS [0]{\spacefactor3000\relax}%
\providecommand \BibitemShut  [1]{\csname bibitem#1\endcsname}%
\let\auto@bib@innerbib\@empty
\bibitem [{\citenamefont {Spohn}(2012)}]{Spohn2012}%
  \BibitemOpen
  \bibfield  {author} {\bibinfo {author} {\bibfnamefont {H.}~\bibnamefont
  {Spohn}},\ }\href@noop {} {\emph {\bibinfo {title} {Large scale dynamics of
  interacting particles}}}\ (\bibinfo  {publisher} {Springer Science \&
  Business Media},\ \bibinfo {year} {2012})\BibitemShut {NoStop}%
\bibitem [{\citenamefont {Birkhoff}(2015)}]{Birkhoff2015}%
  \BibitemOpen
  \bibfield  {author} {\bibinfo {author} {\bibfnamefont {G.}~\bibnamefont
  {Birkhoff}},\ }\href@noop {} {\emph {\bibinfo {title} {Hydrodynamics}}}\
  (\bibinfo  {publisher} {Princeton University Press},\ \bibinfo {year}
  {2015})\BibitemShut {NoStop}%
\bibitem [{\citenamefont {Wyatt}(2006)}]{Wyatt2006}%
  \BibitemOpen
  \bibfield  {author} {\bibinfo {author} {\bibfnamefont {R.~E.}\ \bibnamefont
  {Wyatt}},\ }\href@noop {} {\emph {\bibinfo {title} {Quantum dynamics with
  trajectories: {{Introduction}} to quantum hydrodynamics}}},\ Vol.~\bibinfo
  {volume} {28}\ (\bibinfo  {publisher} {Springer Science \& Business Media},\
  \bibinfo {year} {2006})\BibitemShut {NoStop}%
\bibitem [{\citenamefont {Mukerjee}\ \emph {et~al.}(2006)\citenamefont
  {Mukerjee}, \citenamefont {Oganesyan},\ and\ \citenamefont
  {Huse}}]{Mukerjee2006}%
  \BibitemOpen
  \bibfield  {author} {\bibinfo {author} {\bibfnamefont {S.}~\bibnamefont
  {Mukerjee}}, \bibinfo {author} {\bibfnamefont {V.}~\bibnamefont
  {Oganesyan}},\ and\ \bibinfo {author} {\bibfnamefont {D.}~\bibnamefont
  {Huse}},\ }\bibfield  {title} {\bibinfo {title} {Statistical theory of
  transport by strongly interacting lattice fermions},\ }\href
  {https://doi.org/10.1103/PhysRevB.73.035113} {\bibfield  {journal} {\bibinfo
  {journal} {Phys. Rev. B}\ }\textbf {\bibinfo {volume} {73}},\ \bibinfo
  {pages} {035113} (\bibinfo {year} {2006})}\BibitemShut {NoStop}%
\bibitem [{\citenamefont {Erd{\H{o}}s}\ \emph {et~al.}(2007)\citenamefont
  {Erd{\H{o}}s}, \citenamefont {Schlein},\ and\ \citenamefont
  {Yau}}]{Erdos2007}%
  \BibitemOpen
  \bibfield  {author} {\bibinfo {author} {\bibfnamefont {L.}~\bibnamefont
  {Erd{\H{o}}s}}, \bibinfo {author} {\bibfnamefont {B.}~\bibnamefont
  {Schlein}},\ and\ \bibinfo {author} {\bibfnamefont {H.-T.}\ \bibnamefont
  {Yau}},\ }\bibfield  {title} {\bibinfo {title} {Rigorous derivation of the
  {{Gross}}-{{Pitaevskii}} equation},\ }\href
  {https://doi.org/10.1103/PhysRevLett.98.040404} {\bibfield  {journal}
  {\bibinfo  {journal} {Phys. Rev. Lett.}\ }\textbf {\bibinfo {volume} {98}},\
  \bibinfo {pages} {040404} (\bibinfo {year} {2007})}\BibitemShut {NoStop}%
\bibitem [{\citenamefont {{Castro-Alvaredo}}\ \emph {et~al.}(2016)\citenamefont
  {{Castro-Alvaredo}}, \citenamefont {Doyon},\ and\ \citenamefont
  {Yoshimura}}]{Castro-Alvaredo2016}%
  \BibitemOpen
  \bibfield  {author} {\bibinfo {author} {\bibfnamefont {O.~A.}\ \bibnamefont
  {{Castro-Alvaredo}}}, \bibinfo {author} {\bibfnamefont {B.}~\bibnamefont
  {Doyon}},\ and\ \bibinfo {author} {\bibfnamefont {T.}~\bibnamefont
  {Yoshimura}},\ }\bibfield  {title} {\bibinfo {title} {Emergent hydrodynamics
  in integrable quantum systems out of equilibrium},\ }\href
  {https://doi.org/10.1103/PhysRevX.6.041065} {\bibfield  {journal} {\bibinfo
  {journal} {Phys. Rev. X}\ }\textbf {\bibinfo {volume} {6}},\ \bibinfo {pages}
  {041065} (\bibinfo {year} {2016})}\BibitemShut {NoStop}%
\bibitem [{\citenamefont {Zu}\ \emph {et~al.}(2021)\citenamefont {Zu},
  \citenamefont {Machado}, \citenamefont {Ye}, \citenamefont {Choi},
  \citenamefont {Kobrin}, \citenamefont {Mittiga}, \citenamefont {Hsieh},
  \citenamefont {Bhattacharyya}, \citenamefont {Markham}, \citenamefont
  {Twitchen} \emph {et~al.}}]{Zu2021}%
  \BibitemOpen
  \bibfield  {author} {\bibinfo {author} {\bibfnamefont {C.}~\bibnamefont
  {Zu}}, \bibinfo {author} {\bibfnamefont {F.}~\bibnamefont {Machado}},
  \bibinfo {author} {\bibfnamefont {B.}~\bibnamefont {Ye}}, \bibinfo {author}
  {\bibfnamefont {S.}~\bibnamefont {Choi}}, \bibinfo {author} {\bibfnamefont
  {B.}~\bibnamefont {Kobrin}}, \bibinfo {author} {\bibfnamefont
  {T.}~\bibnamefont {Mittiga}}, \bibinfo {author} {\bibfnamefont
  {S.}~\bibnamefont {Hsieh}}, \bibinfo {author} {\bibfnamefont
  {P.}~\bibnamefont {Bhattacharyya}}, \bibinfo {author} {\bibfnamefont
  {M.}~\bibnamefont {Markham}}, \bibinfo {author} {\bibfnamefont
  {D.}~\bibnamefont {Twitchen}}, \emph {et~al.},\ }\bibfield  {title} {\bibinfo
  {title} {Emergent hydrodynamics in a strongly interacting dipolar spin
  ensemble},\ }\href {https://arxiv.org/abs/2104.07678} {\bibfield  {journal}
  {\bibinfo  {journal} {arXiv:2104.07678}\ } (\bibinfo {year}
  {2021})}\BibitemShut {NoStop}%
\bibitem [{\citenamefont {Kardar}\ \emph {et~al.}(1986)\citenamefont {Kardar},
  \citenamefont {Parisi},\ and\ \citenamefont {Zhang}}]{Kardar1986}%
  \BibitemOpen
  \bibfield  {author} {\bibinfo {author} {\bibfnamefont {M.}~\bibnamefont
  {Kardar}}, \bibinfo {author} {\bibfnamefont {G.}~\bibnamefont {Parisi}},\
  and\ \bibinfo {author} {\bibfnamefont {Y.-C.}\ \bibnamefont {Zhang}},\
  }\bibfield  {title} {\bibinfo {title} {Dynamic scaling of growing
  interfaces},\ }\href {https://doi.org/10.1103/PhysRevLett.56.889} {\bibfield
  {journal} {\bibinfo  {journal} {Phys. Rev. Lett.}\ }\textbf {\bibinfo
  {volume} {56}},\ \bibinfo {pages} {889} (\bibinfo {year} {1986})}\BibitemShut
  {NoStop}%
\bibitem [{\citenamefont {Halpin-Healy}\ and\ \citenamefont
  {Takeuchi}(2015)}]{Halpin2015}%
  \BibitemOpen
  \bibfield  {author} {\bibinfo {author} {\bibfnamefont {T.}~\bibnamefont
  {Halpin-Healy}}\ and\ \bibinfo {author} {\bibfnamefont {K.}~\bibnamefont
  {Takeuchi}},\ }\bibfield  {title} {\bibinfo {title} {A {{KPZ}}
  cocktail-shaken, not stirred...},\ }\href
  {https://doi.org/10.1007/s10955-015-1282-1} {\bibfield  {journal} {\bibinfo
  {journal} {J. Stat. Phys.}\ }\textbf {\bibinfo {volume} {160}},\ \bibinfo
  {pages} {794} (\bibinfo {year} {2015})}\BibitemShut {NoStop}%
\bibitem [{\citenamefont {Spohn}(2020)}]{Spohn2020}%
  \BibitemOpen
  \bibfield  {author} {\bibinfo {author} {\bibfnamefont {H.}~\bibnamefont
  {Spohn}},\ }\bibfield  {title} {\bibinfo {title} {The 1 + 1 dimensional
  {{Kardar}}-{{Parisi}}-{{Zhang}} equation: {{More}} surprises},\ }\href
  {https://doi.org/10.1088/1742-5468/ab712a} {\bibfield  {journal} {\bibinfo
  {journal} {J. Stat. Mech.}\ }\textbf {\bibinfo {volume} {2020}},\ \bibinfo
  {pages} {044001} (\bibinfo {year} {2020})}\BibitemShut {NoStop}%
\bibitem [{\citenamefont {Kadanoff}(1990)}]{Kadanoff1990}%
  \BibitemOpen
  \bibfield  {author} {\bibinfo {author} {\bibfnamefont {L.~P.}\ \bibnamefont
  {Kadanoff}},\ }\bibfield  {title} {\bibinfo {title} {Scaling and universality
  in statistical physics},\ }\href
  {https://doi.org/10.1016/0378-4371(90)90309-G} {\bibfield  {journal}
  {\bibinfo  {journal} {Phys. A: Stat. Mech. Appl.}\ }\textbf {\bibinfo
  {volume} {163}},\ \bibinfo {pages} {1} (\bibinfo {year} {1990})}\BibitemShut
  {NoStop}%
\bibitem [{\citenamefont {Gubinelli}\ and\ \citenamefont
  {Perkowski}(2017)}]{Gubinelli2017}%
  \BibitemOpen
  \bibfield  {author} {\bibinfo {author} {\bibfnamefont {M.}~\bibnamefont
  {Gubinelli}}\ and\ \bibinfo {author} {\bibfnamefont {N.}~\bibnamefont
  {Perkowski}},\ }\bibfield  {title} {\bibinfo {title} {{{KPZ}} reloaded},\
  }\href {https://doi.org/10.1007/s00220-016-2788-3} {\bibfield  {journal}
  {\bibinfo  {journal} {Commun. Math. Phys.}\ }\textbf {\bibinfo {volume}
  {349}},\ \bibinfo {pages} {165} (\bibinfo {year} {2017})}\BibitemShut
  {NoStop}%
\bibitem [{\citenamefont {{\v Z}nidari{\v c}}(2011)}]{Znidaric2011a}%
  \BibitemOpen
  \bibfield  {author} {\bibinfo {author} {\bibfnamefont {M.}~\bibnamefont {{\v
  Z}nidari{\v c}}},\ }\bibfield  {title} {\bibinfo {title} {Spin transport in a
  one-dimensional anisotropic {{Heisenberg}} model},\ }\href
  {https://doi.org/10.1103/PhysRevLett.106.220601} {\bibfield  {journal}
  {\bibinfo  {journal} {Phys. Rev. Lett.}\ }\textbf {\bibinfo {volume} {106}},\
  \bibinfo {pages} {220601} (\bibinfo {year} {2011})}\BibitemShut {NoStop}%
\bibitem [{\citenamefont {Ljubotina}\ \emph {et~al.}(2017)\citenamefont
  {Ljubotina}, \citenamefont {{\v Z}nidari{\v c}},\ and\ \citenamefont
  {Prosen}}]{Ljubotina2017}%
  \BibitemOpen
  \bibfield  {author} {\bibinfo {author} {\bibfnamefont {M.}~\bibnamefont
  {Ljubotina}}, \bibinfo {author} {\bibfnamefont {M.}~\bibnamefont {{\v
  Z}nidari{\v c}}},\ and\ \bibinfo {author} {\bibfnamefont {T.}~\bibnamefont
  {Prosen}},\ }\bibfield  {title} {\bibinfo {title} {Spin diffusion from an
  inhomogeneous quench in an integrable system},\ }\href
  {https://doi.org/10.1038/ncomms16117} {\bibfield  {journal} {\bibinfo
  {journal} {Nat. Commun.}\ }\textbf {\bibinfo {volume} {8}},\ \bibinfo {pages}
  {16117} (\bibinfo {year} {2017})}\BibitemShut {NoStop}%
\bibitem [{\citenamefont {Ljubotina}\ \emph {et~al.}(2019)\citenamefont
  {Ljubotina}, \citenamefont {{\v Z}nidari{\v c}},\ and\ \citenamefont
  {Prosen}}]{Ljubotina2019}%
  \BibitemOpen
  \bibfield  {author} {\bibinfo {author} {\bibfnamefont {M.}~\bibnamefont
  {Ljubotina}}, \bibinfo {author} {\bibfnamefont {M.}~\bibnamefont {{\v
  Z}nidari{\v c}}},\ and\ \bibinfo {author} {\bibfnamefont {T.}~\bibnamefont
  {Prosen}},\ }\bibfield  {title} {\bibinfo {title}
  {Kardar-{{Parisi}}-{{Zhang}} physics in the quantum {{Heisenberg}} magnet},\
  }\href {https://doi.org/10.1103/PhysRevLett.122.210602} {\bibfield  {journal}
  {\bibinfo  {journal} {Phys. Rev. Lett.}\ }\textbf {\bibinfo {volume} {122}},\
  \bibinfo {pages} {210602} (\bibinfo {year} {2019})}\BibitemShut {NoStop}%
\bibitem [{\citenamefont {Gopalakrishnan}\ and\ \citenamefont
  {Vasseur}(2019)}]{Gopalakrishnan2019a}%
  \BibitemOpen
  \bibfield  {author} {\bibinfo {author} {\bibfnamefont {S.}~\bibnamefont
  {Gopalakrishnan}}\ and\ \bibinfo {author} {\bibfnamefont {R.}~\bibnamefont
  {Vasseur}},\ }\bibfield  {title} {\bibinfo {title} {Kinetic theory of spin
  diffusion and superdiffusion in {{XXZ}} spin chains},\ }\href
  {https://doi.org/10.1103/PhysRevLett.122.127202} {\bibfield  {journal}
  {\bibinfo  {journal} {Phys. Rev. Lett.}\ }\textbf {\bibinfo {volume} {122}},\
  \bibinfo {pages} {127202} (\bibinfo {year} {2019})}\BibitemShut {NoStop}%
\bibitem [{\citenamefont {De~Nardis}\ \emph {et~al.}(2019)\citenamefont
  {De~Nardis}, \citenamefont {Medenjak}, \citenamefont {Karrasch},\ and\
  \citenamefont {Ilievski}}]{DeNardis2019}%
  \BibitemOpen
  \bibfield  {author} {\bibinfo {author} {\bibfnamefont {J.}~\bibnamefont
  {De~Nardis}}, \bibinfo {author} {\bibfnamefont {M.}~\bibnamefont {Medenjak}},
  \bibinfo {author} {\bibfnamefont {C.}~\bibnamefont {Karrasch}},\ and\
  \bibinfo {author} {\bibfnamefont {E.}~\bibnamefont {Ilievski}},\ }\bibfield
  {title} {\bibinfo {title} {Anomalous spin diffusion in one-dimensional
  antiferromagnets},\ }\href {https://doi.org/10.1103/PhysRevLett.123.186601}
  {\bibfield  {journal} {\bibinfo  {journal} {Phys. Rev. Lett.}\ }\textbf
  {\bibinfo {volume} {123}},\ \bibinfo {pages} {186601} (\bibinfo {year}
  {2019})}\BibitemShut {NoStop}%
\bibitem [{\citenamefont {Gopalakrishnan}\ \emph {et~al.}(2019)\citenamefont
  {Gopalakrishnan}, \citenamefont {Vasseur},\ and\ \citenamefont
  {Ware}}]{Gopalakrishnan2019}%
  \BibitemOpen
  \bibfield  {author} {\bibinfo {author} {\bibfnamefont {S.}~\bibnamefont
  {Gopalakrishnan}}, \bibinfo {author} {\bibfnamefont {R.}~\bibnamefont
  {Vasseur}},\ and\ \bibinfo {author} {\bibfnamefont {B.}~\bibnamefont
  {Ware}},\ }\bibfield  {title} {\bibinfo {title} {Anomalous relaxation and the
  high-temperature structure factor of {{XXZ}} spin chains},\ }\href
  {https://doi.org/10.1073/pnas.1906914116} {\bibfield  {journal} {\bibinfo
  {journal} {PNAS}\ }\textbf {\bibinfo {volume} {116}},\ \bibinfo {pages}
  {16250} (\bibinfo {year} {2019})}\BibitemShut {NoStop}%
\bibitem [{\citenamefont {Bulchandani}(2020)}]{Bulchandani2020a}%
  \BibitemOpen
  \bibfield  {author} {\bibinfo {author} {\bibfnamefont {V.~B.}\ \bibnamefont
  {Bulchandani}},\ }\bibfield  {title} {\bibinfo {title}
  {Kardar-{{Parisi}}-{{Zhang}} universality from soft gauge modes},\ }\href
  {https://doi.org/10.1103/PhysRevB.101.041411} {\bibfield  {journal} {\bibinfo
   {journal} {Phys. Rev. B}\ }\textbf {\bibinfo {volume} {101}},\ \bibinfo
  {pages} {041411} (\bibinfo {year} {2020})}\BibitemShut {NoStop}%
\bibitem [{\citenamefont {Bertini}\ \emph {et~al.}(2016)\citenamefont
  {Bertini}, \citenamefont {Collura}, \citenamefont {De~Nardis},\ and\
  \citenamefont {Fagotti}}]{Bertini2016}%
  \BibitemOpen
  \bibfield  {author} {\bibinfo {author} {\bibfnamefont {B.}~\bibnamefont
  {Bertini}}, \bibinfo {author} {\bibfnamefont {M.}~\bibnamefont {Collura}},
  \bibinfo {author} {\bibfnamefont {J.}~\bibnamefont {De~Nardis}},\ and\
  \bibinfo {author} {\bibfnamefont {M.}~\bibnamefont {Fagotti}},\ }\bibfield
  {title} {\bibinfo {title} {Transport in out-of-equilibrium {{XXZ}} chains:
  {{Exact}} profiles of charges and currents},\ }\href
  {https://doi.org/10.1103/PhysRevLett.117.207201} {\bibfield  {journal}
  {\bibinfo  {journal} {Phys. Rev. Lett.}\ }\textbf {\bibinfo {volume} {117}},\
  \bibinfo {pages} {207201} (\bibinfo {year} {2016})}\BibitemShut {NoStop}%
\bibitem [{\citenamefont {Ilievski}\ and\ \citenamefont
  {De~Nardis}(2017)}]{Ilievski2017}%
  \BibitemOpen
  \bibfield  {author} {\bibinfo {author} {\bibfnamefont {E.}~\bibnamefont
  {Ilievski}}\ and\ \bibinfo {author} {\bibfnamefont {J.}~\bibnamefont
  {De~Nardis}},\ }\bibfield  {title} {\bibinfo {title} {Microscopic origin of
  ideal conductivity in integrable quantum models},\ }\href
  {https://doi.org/10.1103/PhysRevLett.119.020602} {\bibfield  {journal}
  {\bibinfo  {journal} {Phys. Rev. Lett.}\ }\textbf {\bibinfo {volume} {119}},\
  \bibinfo {pages} {020602} (\bibinfo {year} {2017})}\BibitemShut {NoStop}%
\bibitem [{\citenamefont {Bulchandani}\ \emph {et~al.}(2018)\citenamefont
  {Bulchandani}, \citenamefont {Vasseur}, \citenamefont {Karrasch},\ and\
  \citenamefont {Moore}}]{Bulchandani2018}%
  \BibitemOpen
  \bibfield  {author} {\bibinfo {author} {\bibfnamefont {V.~B.}\ \bibnamefont
  {Bulchandani}}, \bibinfo {author} {\bibfnamefont {R.}~\bibnamefont
  {Vasseur}}, \bibinfo {author} {\bibfnamefont {C.}~\bibnamefont {Karrasch}},\
  and\ \bibinfo {author} {\bibfnamefont {J.~E.}\ \bibnamefont {Moore}},\
  }\bibfield  {title} {\bibinfo {title} {Bethe-{{Boltzmann}} hydrodynamics and
  spin transport in the {{XXZ}} chain},\ }\href
  {https://doi.org/10.1103/PhysRevB.97.045407} {\bibfield  {journal} {\bibinfo
  {journal} {Phys. Rev. B}\ }\textbf {\bibinfo {volume} {97}},\ \bibinfo
  {pages} {045407} (\bibinfo {year} {2018})}\BibitemShut {NoStop}%
\bibitem [{\citenamefont {Prosen}\ and\ \citenamefont {{\v Z}unkovi{\v
  c}}(2013)}]{Prosen2013}%
  \BibitemOpen
  \bibfield  {author} {\bibinfo {author} {\bibfnamefont {T.}~\bibnamefont
  {Prosen}}\ and\ \bibinfo {author} {\bibfnamefont {B.}~\bibnamefont {{\v
  Z}unkovi{\v c}}},\ }\bibfield  {title} {\bibinfo {title} {Macroscopic
  diffusive transport in a microscopically integrable {{Hamiltonian}} system},\
  }\href {https://doi.org/10.1103/PhysRevLett.111.040602} {\bibfield  {journal}
  {\bibinfo  {journal} {Phys. Rev. Lett.}\ }\textbf {\bibinfo {volume} {111}},\
  \bibinfo {pages} {040602} (\bibinfo {year} {2013})}\BibitemShut {NoStop}%
\bibitem [{\citenamefont {Das}\ \emph {et~al.}(2019)\citenamefont {Das},
  \citenamefont {Kulkarni}, \citenamefont {Spohn},\ and\ \citenamefont
  {Dhar}}]{Das2019}%
  \BibitemOpen
  \bibfield  {author} {\bibinfo {author} {\bibfnamefont {A.}~\bibnamefont
  {Das}}, \bibinfo {author} {\bibfnamefont {M.}~\bibnamefont {Kulkarni}},
  \bibinfo {author} {\bibfnamefont {H.}~\bibnamefont {Spohn}},\ and\ \bibinfo
  {author} {\bibfnamefont {A.}~\bibnamefont {Dhar}},\ }\bibfield  {title}
  {\bibinfo {title} {Kardar-{{Parisi}}-{{Zhang}} scaling for an integrable
  lattice {{Landau}}-{{Lifshitz}} spin chain},\ }\href
  {https://doi.org/10.1103/PhysRevE.100.042116} {\bibfield  {journal} {\bibinfo
   {journal} {Phys. Rev. E}\ }\textbf {\bibinfo {volume} {100}},\ \bibinfo
  {pages} {042116} (\bibinfo {year} {2019})}\BibitemShut {NoStop}%
\bibitem [{\citenamefont {Krajnik}\ and\ \citenamefont
  {Prosen}(2020)}]{Krajnik2020}%
  \BibitemOpen
  \bibfield  {author} {\bibinfo {author} {\bibfnamefont {{\v Z}.}~\bibnamefont
  {Krajnik}}\ and\ \bibinfo {author} {\bibfnamefont {T.}~\bibnamefont
  {Prosen}},\ }\bibfield  {title} {\bibinfo {title}
  {Kardar-{{Parisi}}-{{Zhang}} physics in integrable rotationally symmetric
  dynamics on discrete space-time lattice},\ }\href
  {https://doi.org/10.1007/s10955-020-02523-1} {\bibfield  {journal} {\bibinfo
  {journal} {J. Stat. Phys.}\ }\textbf {\bibinfo {volume} {179}},\ \bibinfo
  {pages} {110} (\bibinfo {year} {2020})}\BibitemShut {NoStop}%
\bibitem [{\citenamefont {Krajnik}\ \emph {et~al.}(2020)\citenamefont
  {Krajnik}, \citenamefont {Ilievski},\ and\ \citenamefont
  {Prosen}}]{Krajnik2020a}%
  \BibitemOpen
  \bibfield  {author} {\bibinfo {author} {\bibfnamefont {{\v Z}.}~\bibnamefont
  {Krajnik}}, \bibinfo {author} {\bibfnamefont {E.}~\bibnamefont {Ilievski}},\
  and\ \bibinfo {author} {\bibfnamefont {T.}~\bibnamefont {Prosen}},\
  }\bibfield  {title} {\bibinfo {title} {Integrable matrix models in discrete
  space-time},\ }\href {https://doi.org/10.21468/SciPostPhys.9.3.038}
  {\bibfield  {journal} {\bibinfo  {journal} {SciPost Phys.}\ }\textbf
  {\bibinfo {volume} {9}},\ \bibinfo {pages} {38} (\bibinfo {year}
  {2020})}\BibitemShut {NoStop}%
\bibitem [{\citenamefont {Ilievski}\ \emph {et~al.}(2020)\citenamefont
  {Ilievski}, \citenamefont {De~Nardis}, \citenamefont {Gopalakrishnan},
  \citenamefont {Vasseur},\ and\ \citenamefont {Ware}}]{Ilievski2020}%
  \BibitemOpen
  \bibfield  {author} {\bibinfo {author} {\bibfnamefont {E.}~\bibnamefont
  {Ilievski}}, \bibinfo {author} {\bibfnamefont {J.}~\bibnamefont {De~Nardis}},
  \bibinfo {author} {\bibfnamefont {S.}~\bibnamefont {Gopalakrishnan}},
  \bibinfo {author} {\bibfnamefont {R.}~\bibnamefont {Vasseur}},\ and\ \bibinfo
  {author} {\bibfnamefont {B.}~\bibnamefont {Ware}},\ }\bibfield  {title}
  {\bibinfo {title} {Superuniversality of superdiffusion},\ }\href
  {https://arxiv.org/abs/2009.08425} {\bibfield  {journal} {\bibinfo  {journal}
  {arXiv:2009.08425}\ } (\bibinfo {year} {2020})}\BibitemShut {NoStop}%
\bibitem [{\citenamefont {Bertini}\ \emph {et~al.}(2021)\citenamefont
  {Bertini}, \citenamefont {Heidrich-Meisner}, \citenamefont {Karrasch},
  \citenamefont {Prosen}, \citenamefont {Steinigeweg},\ and\ \citenamefont
  {{\v{Z}}nidari{\v{c}}}}]{Bertini2021}%
  \BibitemOpen
  \bibfield  {author} {\bibinfo {author} {\bibfnamefont {B.}~\bibnamefont
  {Bertini}}, \bibinfo {author} {\bibfnamefont {F.}~\bibnamefont
  {Heidrich-Meisner}}, \bibinfo {author} {\bibfnamefont {C.}~\bibnamefont
  {Karrasch}}, \bibinfo {author} {\bibfnamefont {T.}~\bibnamefont {Prosen}},
  \bibinfo {author} {\bibfnamefont {R.}~\bibnamefont {Steinigeweg}},\ and\
  \bibinfo {author} {\bibfnamefont {M.}~\bibnamefont {{\v{Z}}nidari{\v{c}}}},\
  }\bibfield  {title} {\bibinfo {title} {Finite-temperature transport in
  one-dimensional quantum lattice models},\ }\href
  {https://doi.org/10.1103/RevModPhys.93.025003} {\bibfield  {journal}
  {\bibinfo  {journal} {Rev. Mod. Phys.}\ }\textbf {\bibinfo {volume} {93}},\
  \bibinfo {pages} {025003} (\bibinfo {year} {2021})}\BibitemShut {NoStop}%
\bibitem [{\citenamefont {Bulchandani}\ \emph {et~al.}(2021)\citenamefont
  {Bulchandani}, \citenamefont {Gopalakrishnan},\ and\ \citenamefont
  {Ilievski}}]{Bulchandani2021}%
  \BibitemOpen
  \bibfield  {author} {\bibinfo {author} {\bibfnamefont {V.~B.}\ \bibnamefont
  {Bulchandani}}, \bibinfo {author} {\bibfnamefont {S.}~\bibnamefont
  {Gopalakrishnan}},\ and\ \bibinfo {author} {\bibfnamefont {E.}~\bibnamefont
  {Ilievski}},\ }\bibfield  {title} {\bibinfo {title} {Superdiffusion in spin
  chains},\ }\href {https://arxiv.org/abs/2103.01976} {\bibfield  {journal}
  {\bibinfo  {journal} {arXiv:2103.01976}\ } (\bibinfo {year}
  {2021})}\BibitemShut {NoStop}%
\bibitem [{\citenamefont {Hild}\ \emph {et~al.}(2014)\citenamefont {Hild},
  \citenamefont {Fukuhara}, \citenamefont {Schau{\ss}}, \citenamefont {Zeiher},
  \citenamefont {Knap}, \citenamefont {Demler}, \citenamefont {Bloch},\ and\
  \citenamefont {Gross}}]{Hild2014}%
  \BibitemOpen
  \bibfield  {author} {\bibinfo {author} {\bibfnamefont {S.}~\bibnamefont
  {Hild}}, \bibinfo {author} {\bibfnamefont {T.}~\bibnamefont {Fukuhara}},
  \bibinfo {author} {\bibfnamefont {P.}~\bibnamefont {Schau{\ss}}}, \bibinfo
  {author} {\bibfnamefont {J.}~\bibnamefont {Zeiher}}, \bibinfo {author}
  {\bibfnamefont {M.}~\bibnamefont {Knap}}, \bibinfo {author} {\bibfnamefont
  {E.}~\bibnamefont {Demler}}, \bibinfo {author} {\bibfnamefont
  {I.}~\bibnamefont {Bloch}},\ and\ \bibinfo {author} {\bibfnamefont
  {C.}~\bibnamefont {Gross}},\ }\bibfield  {title} {\bibinfo {title}
  {Far-from-equilibrium spin transport in {{Heisenberg}} quantum magnets},\
  }\href {https://doi.org/10.1103/PhysRevLett.113.147205} {\bibfield  {journal}
  {\bibinfo  {journal} {Phys. Rev. Lett.}\ }\textbf {\bibinfo {volume} {113}},\
  \bibinfo {pages} {1} (\bibinfo {year} {2014})}\BibitemShut {NoStop}%
\bibitem [{\citenamefont {Jepsen}\ \emph {et~al.}(2020)\citenamefont {Jepsen},
  \citenamefont {{Amato-Grill}}, \citenamefont {Dimitrova}, \citenamefont {Ho},
  \citenamefont {Demler},\ and\ \citenamefont {Ketterle}}]{Jepsen2020}%
  \BibitemOpen
  \bibfield  {author} {\bibinfo {author} {\bibfnamefont {P.~N.}\ \bibnamefont
  {Jepsen}}, \bibinfo {author} {\bibfnamefont {J.}~\bibnamefont
  {{Amato-Grill}}}, \bibinfo {author} {\bibfnamefont {I.}~\bibnamefont
  {Dimitrova}}, \bibinfo {author} {\bibfnamefont {W.~W.}\ \bibnamefont {Ho}},
  \bibinfo {author} {\bibfnamefont {E.}~\bibnamefont {Demler}},\ and\ \bibinfo
  {author} {\bibfnamefont {W.}~\bibnamefont {Ketterle}},\ }\bibfield  {title}
  {\bibinfo {title} {Spin transport in a tunable {{Heisenberg}} model realized
  with ultracold atoms},\ }\href {https://doi.org/10.1038/s41586-020-3033-y}
  {\bibfield  {journal} {\bibinfo  {journal} {Nature}\ }\textbf {\bibinfo
  {volume} {588}},\ \bibinfo {pages} {403} (\bibinfo {year}
  {2020})}\BibitemShut {NoStop}%
\bibitem [{\citenamefont {Jepsen}\ \emph {et~al.}(2021)\citenamefont {Jepsen},
  \citenamefont {Ho}, \citenamefont {{Amato-Grill}}, \citenamefont {Dimitrova},
  \citenamefont {Demler},\ and\ \citenamefont {Ketterle}}]{Jepsen2021}%
  \BibitemOpen
  \bibfield  {author} {\bibinfo {author} {\bibfnamefont {P.~N.}\ \bibnamefont
  {Jepsen}}, \bibinfo {author} {\bibfnamefont {W.~W.}\ \bibnamefont {Ho}},
  \bibinfo {author} {\bibfnamefont {J.}~\bibnamefont {{Amato-Grill}}}, \bibinfo
  {author} {\bibfnamefont {I.}~\bibnamefont {Dimitrova}}, \bibinfo {author}
  {\bibfnamefont {E.}~\bibnamefont {Demler}},\ and\ \bibinfo {author}
  {\bibfnamefont {W.}~\bibnamefont {Ketterle}},\ }\bibfield  {title} {\bibinfo
  {title} {Transverse spin dynamics in the anisotropic {{Heisenberg}} model
  realized with ultracold atoms},\ }\href {https://arxiv.org/abs/2103.07866}
  {\bibfield  {journal} {\bibinfo  {journal} {arXiv:2103.07866}\ } (\bibinfo
  {year} {2021})}\BibitemShut {NoStop}%
\bibitem [{\citenamefont {Scheie}\ \emph {et~al.}(2021)\citenamefont {Scheie},
  \citenamefont {Sherman}, \citenamefont {Dupont}, \citenamefont {Nagler},
  \citenamefont {Stone}, \citenamefont {Granroth}, \citenamefont {Moore},\ and\
  \citenamefont {Tennant}}]{Scheie2021}%
  \BibitemOpen
  \bibfield  {author} {\bibinfo {author} {\bibfnamefont {A.}~\bibnamefont
  {Scheie}}, \bibinfo {author} {\bibfnamefont {N.~E.}\ \bibnamefont {Sherman}},
  \bibinfo {author} {\bibfnamefont {M.}~\bibnamefont {Dupont}}, \bibinfo
  {author} {\bibfnamefont {S.~E.}\ \bibnamefont {Nagler}}, \bibinfo {author}
  {\bibfnamefont {M.~B.}\ \bibnamefont {Stone}}, \bibinfo {author}
  {\bibfnamefont {G.~E.}\ \bibnamefont {Granroth}}, \bibinfo {author}
  {\bibfnamefont {J.~E.}\ \bibnamefont {Moore}},\ and\ \bibinfo {author}
  {\bibfnamefont {D.~A.}\ \bibnamefont {Tennant}},\ }\bibfield  {title}
  {\bibinfo {title} {Detection of {{Kardar}}-{{Parisi}}-{{Zhang}} hydrodynamics
  in a quantum {{Heisenberg}} spin-1/2 chain},\ }\href
  {https://doi.org/10.1038/s41567-021-01191-6} {\bibfield  {journal} {\bibinfo
  {journal} {Nat. Phys.}\ }\textbf {\bibinfo {volume} {17}},\ \bibinfo {pages}
  {726} (\bibinfo {year} {2021})}\BibitemShut {NoStop}%
\bibitem [{\citenamefont {Hartmann}\ \emph {et~al.}(2018)\citenamefont
  {Hartmann}, \citenamefont {Le~Doussal}, \citenamefont {Majumdar},
  \citenamefont {Rosso},\ and\ \citenamefont {Schehr}}]{Hartmann2018}%
  \BibitemOpen
  \bibfield  {author} {\bibinfo {author} {\bibfnamefont {A.~K.}\ \bibnamefont
  {Hartmann}}, \bibinfo {author} {\bibfnamefont {P.}~\bibnamefont
  {Le~Doussal}}, \bibinfo {author} {\bibfnamefont {S.~N.}\ \bibnamefont
  {Majumdar}}, \bibinfo {author} {\bibfnamefont {A.}~\bibnamefont {Rosso}},\
  and\ \bibinfo {author} {\bibfnamefont {G.}~\bibnamefont {Schehr}},\
  }\bibfield  {title} {\bibinfo {title} {High-precision simulation of the
  height distribution for the {{KPZ}} equation},\ }\href
  {https://doi.org/10.1209/0295-5075/121/67004} {\bibfield  {journal} {\bibinfo
   {journal} {EPL}\ }\textbf {\bibinfo {volume} {121}},\ \bibinfo {pages}
  {67004} (\bibinfo {year} {2018})}\BibitemShut {NoStop}%
\bibitem [{\citenamefont {Duan}\ \emph {et~al.}(2003)\citenamefont {Duan},
  \citenamefont {Demler},\ and\ \citenamefont {Lukin}}]{Duan2003}%
  \BibitemOpen
  \bibfield  {author} {\bibinfo {author} {\bibfnamefont {L.-M.}\ \bibnamefont
  {Duan}}, \bibinfo {author} {\bibfnamefont {E.}~\bibnamefont {Demler}},\ and\
  \bibinfo {author} {\bibfnamefont {M.~D.}\ \bibnamefont {Lukin}},\ }\bibfield
  {title} {\bibinfo {title} {Controlling spin exchange interactions of
  ultracold atoms in optical lattices},\ }\href
  {https://doi.org/10.1103/PhysRevLett.91.090402} {\bibfield  {journal}
  {\bibinfo  {journal} {Phys. Rev. Lett.}\ }\textbf {\bibinfo {volume} {91}},\
  \bibinfo {pages} {090402} (\bibinfo {year} {2003})}\BibitemShut {NoStop}%
\bibitem [{\citenamefont {Kuklov}\ and\ \citenamefont
  {Svistunov}(2003)}]{Kuklov2003}%
  \BibitemOpen
  \bibfield  {author} {\bibinfo {author} {\bibfnamefont {A.~B.}\ \bibnamefont
  {Kuklov}}\ and\ \bibinfo {author} {\bibfnamefont {B.~V.}\ \bibnamefont
  {Svistunov}},\ }\bibfield  {title} {\bibinfo {title} {Counterflow
  superfluidity of two-species ultracold atoms in a commensurate optical
  lattice},\ }\href {https://doi.org/10.1103/PhysRevLett.90.100401} {\bibfield
  {journal} {\bibinfo  {journal} {Phys. Rev. Lett.}\ }\textbf {\bibinfo
  {volume} {90}},\ \bibinfo {pages} {100401} (\bibinfo {year}
  {2003})}\BibitemShut {NoStop}%
\bibitem [{SI()}]{SI}%
  \BibitemOpen
  \href@noop {} {}\bibinfo {note} {{see Supplementary Information}}\BibitemShut
  {NoStop}%
\bibitem [{\citenamefont {Fukuhara}\ \emph {et~al.}(2013)\citenamefont
  {Fukuhara}, \citenamefont {Kantian}, \citenamefont {Endres}, \citenamefont
  {Cheneau}, \citenamefont {Schau{\ss}}, \citenamefont {Hild}, \citenamefont
  {Bellem}, \citenamefont {Schollw{\"o}ck}, \citenamefont {Giamarchi},
  \citenamefont {Gross}, \citenamefont {Bloch},\ and\ \citenamefont
  {Kuhr}}]{Fukuhara2013}%
  \BibitemOpen
  \bibfield  {author} {\bibinfo {author} {\bibfnamefont {T.}~\bibnamefont
  {Fukuhara}}, \bibinfo {author} {\bibfnamefont {A.}~\bibnamefont {Kantian}},
  \bibinfo {author} {\bibfnamefont {M.}~\bibnamefont {Endres}}, \bibinfo
  {author} {\bibfnamefont {M.}~\bibnamefont {Cheneau}}, \bibinfo {author}
  {\bibfnamefont {P.}~\bibnamefont {Schau{\ss}}}, \bibinfo {author}
  {\bibfnamefont {S.}~\bibnamefont {Hild}}, \bibinfo {author} {\bibfnamefont
  {D.}~\bibnamefont {Bellem}}, \bibinfo {author} {\bibfnamefont
  {U.}~\bibnamefont {Schollw{\"o}ck}}, \bibinfo {author} {\bibfnamefont
  {T.}~\bibnamefont {Giamarchi}}, \bibinfo {author} {\bibfnamefont
  {C.}~\bibnamefont {Gross}}, \bibinfo {author} {\bibfnamefont
  {I.}~\bibnamefont {Bloch}},\ and\ \bibinfo {author} {\bibfnamefont
  {S.}~\bibnamefont {Kuhr}},\ }\bibfield  {title} {\bibinfo {title} {Quantum
  dynamics of a mobile spin impurity},\ }\href
  {https://doi.org/10.1038/nphys2561} {\bibfield  {journal} {\bibinfo
  {journal} {Nat. Phys.}\ }\textbf {\bibinfo {volume} {9}},\ \bibinfo {pages}
  {235} (\bibinfo {year} {2013})}\BibitemShut {NoStop}%
\bibitem [{\citenamefont {Halimeh}\ \emph {et~al.}(2014)\citenamefont
  {Halimeh}, \citenamefont {W{\"o}llert}, \citenamefont {McCulloch},
  \citenamefont {Schollw{\"o}ck},\ and\ \citenamefont {Barthel}}]{Halimeh2014}%
  \BibitemOpen
  \bibfield  {author} {\bibinfo {author} {\bibfnamefont {J.~C.}\ \bibnamefont
  {Halimeh}}, \bibinfo {author} {\bibfnamefont {A.}~\bibnamefont
  {W{\"o}llert}}, \bibinfo {author} {\bibfnamefont {I.}~\bibnamefont
  {McCulloch}}, \bibinfo {author} {\bibfnamefont {U.}~\bibnamefont
  {Schollw{\"o}ck}},\ and\ \bibinfo {author} {\bibfnamefont {T.}~\bibnamefont
  {Barthel}},\ }\bibfield  {title} {\bibinfo {title} {Domain-wall melting in
  ultracold-boson systems with hole and spin-flip defects},\ }\href
  {https://doi.org/10.1103/PhysRevA.89.063603} {\bibfield  {journal} {\bibinfo
  {journal} {Phys. Rev. A}\ }\textbf {\bibinfo {volume} {89}},\ \bibinfo
  {pages} {063603} (\bibinfo {year} {2014})}\BibitemShut {NoStop}%
\bibitem [{\citenamefont {Misguich}\ \emph {et~al.}(2017)\citenamefont
  {Misguich}, \citenamefont {Mallick},\ and\ \citenamefont
  {Krapivsky}}]{Misguich2017}%
  \BibitemOpen
  \bibfield  {author} {\bibinfo {author} {\bibfnamefont {G.}~\bibnamefont
  {Misguich}}, \bibinfo {author} {\bibfnamefont {K.}~\bibnamefont {Mallick}},\
  and\ \bibinfo {author} {\bibfnamefont {P.~L.}\ \bibnamefont {Krapivsky}},\
  }\bibfield  {title} {\bibinfo {title} {Dynamics of the spin-1/2
  {{Heisenberg}} chain initialized in a domain-wall state},\ }\href
  {https://doi.org/10.1103/PhysRevB.96.195151} {\bibfield  {journal} {\bibinfo
  {journal} {Phys. Rev. B}\ }\textbf {\bibinfo {volume} {96}},\ \bibinfo
  {pages} {195151} (\bibinfo {year} {2017})}\BibitemShut {NoStop}%
\bibitem [{\citenamefont {White}\ \emph {et~al.}(2018)\citenamefont {White},
  \citenamefont {Zaletel}, \citenamefont {Mong},\ and\ \citenamefont
  {Refael}}]{White2018}%
  \BibitemOpen
  \bibfield  {author} {\bibinfo {author} {\bibfnamefont {C.~D.}\ \bibnamefont
  {White}}, \bibinfo {author} {\bibfnamefont {M.}~\bibnamefont {Zaletel}},
  \bibinfo {author} {\bibfnamefont {R.~S.~K.}\ \bibnamefont {Mong}},\ and\
  \bibinfo {author} {\bibfnamefont {G.}~\bibnamefont {Refael}},\ }\bibfield
  {title} {\bibinfo {title} {Quantum dynamics of thermalizing systems},\ }\href
  {https://doi.org/10.1103/PhysRevB.97.035127} {\bibfield  {journal} {\bibinfo
  {journal} {Phys. Rev. B}\ }\textbf {\bibinfo {volume} {97}},\ \bibinfo
  {pages} {035127} (\bibinfo {year} {2018})}\BibitemShut {NoStop}%
\bibitem [{\citenamefont {Ye}\ \emph {et~al.}(2020)\citenamefont {Ye},
  \citenamefont {Machado}, \citenamefont {White}, \citenamefont {Mong},\ and\
  \citenamefont {Yao}}]{Ye2020}%
  \BibitemOpen
  \bibfield  {author} {\bibinfo {author} {\bibfnamefont {B.}~\bibnamefont
  {Ye}}, \bibinfo {author} {\bibfnamefont {F.}~\bibnamefont {Machado}},
  \bibinfo {author} {\bibfnamefont {C.~D.}\ \bibnamefont {White}}, \bibinfo
  {author} {\bibfnamefont {R.~S.~K.}\ \bibnamefont {Mong}},\ and\ \bibinfo
  {author} {\bibfnamefont {N.~Y.}\ \bibnamefont {Yao}},\ }\bibfield  {title}
  {\bibinfo {title} {Emergent hydrodynamics in nonequilibrium quantum
  systems},\ }\href {https://doi.org/10.1103/PhysRevLett.125.030601} {\bibfield
   {journal} {\bibinfo  {journal} {Phys. Rev. Lett.}\ }\textbf {\bibinfo
  {volume} {125}},\ \bibinfo {pages} {030601} (\bibinfo {year}
  {2020})}\BibitemShut {NoStop}%
\bibitem [{\citenamefont {Gamayun}\ \emph {et~al.}(2019)\citenamefont
  {Gamayun}, \citenamefont {Miao},\ and\ \citenamefont
  {Ilievski}}]{Gamayun2019}%
  \BibitemOpen
  \bibfield  {author} {\bibinfo {author} {\bibfnamefont {O.}~\bibnamefont
  {Gamayun}}, \bibinfo {author} {\bibfnamefont {Y.}~\bibnamefont {Miao}},\ and\
  \bibinfo {author} {\bibfnamefont {E.}~\bibnamefont {Ilievski}},\ }\bibfield
  {title} {\bibinfo {title} {Domain-wall dynamics in the
  {{Landau}}-{{Lifshitz}} magnet and the classical-quantum correspondence for
  spin transport},\ }\href {https://doi.org/10.1103/PhysRevB.99.140301}
  {\bibfield  {journal} {\bibinfo  {journal} {Phys. Rev. B}\ }\textbf {\bibinfo
  {volume} {99}},\ \bibinfo {pages} {140301} (\bibinfo {year}
  {2019})}\BibitemShut {NoStop}%
\bibitem [{\citenamefont {De~Nardis}\ \emph {et~al.}(2020)\citenamefont
  {De~Nardis}, \citenamefont {Gopalakrishnan}, \citenamefont {Ilievski},\ and\
  \citenamefont {Vasseur}}]{DeNardis2020}%
  \BibitemOpen
  \bibfield  {author} {\bibinfo {author} {\bibfnamefont {J.}~\bibnamefont
  {De~Nardis}}, \bibinfo {author} {\bibfnamefont {S.}~\bibnamefont
  {Gopalakrishnan}}, \bibinfo {author} {\bibfnamefont {E.}~\bibnamefont
  {Ilievski}},\ and\ \bibinfo {author} {\bibfnamefont {R.}~\bibnamefont
  {Vasseur}},\ }\bibfield  {title} {\bibinfo {title} {Superdiffusion from
  emergent classical solitons in quantum spin chains},\ }\href
  {https://doi.org/10.1103/PhysRevLett.125.070601} {\bibfield  {journal}
  {\bibinfo  {journal} {Phys. Rev. Lett.}\ }\textbf {\bibinfo {volume} {125}},\
  \bibinfo {pages} {070601} (\bibinfo {year} {2020})}\BibitemShut {NoStop}%
\bibitem [{\citenamefont {Tang}\ \emph {et~al.}(2018)\citenamefont {Tang},
  \citenamefont {Kao}, \citenamefont {Li}, \citenamefont {Seo}, \citenamefont
  {Mallayya}, \citenamefont {Rigol}, \citenamefont {Gopalakrishnan},\ and\
  \citenamefont {Lev}}]{Tang2018}%
  \BibitemOpen
  \bibfield  {author} {\bibinfo {author} {\bibfnamefont {Y.}~\bibnamefont
  {Tang}}, \bibinfo {author} {\bibfnamefont {W.}~\bibnamefont {Kao}}, \bibinfo
  {author} {\bibfnamefont {K.-Y.}\ \bibnamefont {Li}}, \bibinfo {author}
  {\bibfnamefont {S.}~\bibnamefont {Seo}}, \bibinfo {author} {\bibfnamefont
  {K.}~\bibnamefont {Mallayya}}, \bibinfo {author} {\bibfnamefont
  {M.}~\bibnamefont {Rigol}}, \bibinfo {author} {\bibfnamefont
  {S.}~\bibnamefont {Gopalakrishnan}},\ and\ \bibinfo {author} {\bibfnamefont
  {B.~L.}\ \bibnamefont {Lev}},\ }\bibfield  {title} {\bibinfo {title}
  {Thermalization near integrability in a dipolar quantum {{Newton}}'s
  cradle},\ }\href {https://doi.org/10.1103/PhysRevX.8.021030} {\bibfield
  {journal} {\bibinfo  {journal} {Phys. Rev. X}\ }\textbf {\bibinfo {volume}
  {8}},\ \bibinfo {pages} {021030} (\bibinfo {year} {2018})}\BibitemShut
  {NoStop}%
\bibitem [{\citenamefont {Nichols}\ \emph {et~al.}(2019)\citenamefont
  {Nichols}, \citenamefont {Cheuk}, \citenamefont {Okan}, \citenamefont
  {Hartke}, \citenamefont {Mendez}, \citenamefont {Senthil}, \citenamefont
  {Khatami}, \citenamefont {Zhang},\ and\ \citenamefont
  {Zwierlein}}]{Nichols2019}%
  \BibitemOpen
  \bibfield  {author} {\bibinfo {author} {\bibfnamefont {M.~A.}\ \bibnamefont
  {Nichols}}, \bibinfo {author} {\bibfnamefont {L.~W.}\ \bibnamefont {Cheuk}},
  \bibinfo {author} {\bibfnamefont {M.}~\bibnamefont {Okan}}, \bibinfo {author}
  {\bibfnamefont {T.~R.}\ \bibnamefont {Hartke}}, \bibinfo {author}
  {\bibfnamefont {E.}~\bibnamefont {Mendez}}, \bibinfo {author} {\bibfnamefont
  {T.}~\bibnamefont {Senthil}}, \bibinfo {author} {\bibfnamefont
  {E.}~\bibnamefont {Khatami}}, \bibinfo {author} {\bibfnamefont
  {H.}~\bibnamefont {Zhang}},\ and\ \bibinfo {author} {\bibfnamefont {M.~W.}\
  \bibnamefont {Zwierlein}},\ }\bibfield  {title} {\bibinfo {title} {Spin
  transport in a {{Mott}} insulator of ultracold fermions},\ }\href
  {https://doi.org/10.1126/science.aat4387} {\bibfield  {journal} {\bibinfo
  {journal} {Science}\ }\textbf {\bibinfo {volume} {363}},\ \bibinfo {pages}
  {383} (\bibinfo {year} {2019})}\BibitemShut {NoStop}%
\bibitem [{\citenamefont {De~Nardis}\ \emph {et~al.}(2021)\citenamefont
  {De~Nardis}, \citenamefont {Gopalakrishnan}, \citenamefont {Vasseur},\ and\
  \citenamefont {Ware}}]{DeNardis2021}%
  \BibitemOpen
  \bibfield  {author} {\bibinfo {author} {\bibfnamefont {J.}~\bibnamefont
  {De~Nardis}}, \bibinfo {author} {\bibfnamefont {S.}~\bibnamefont
  {Gopalakrishnan}}, \bibinfo {author} {\bibfnamefont {R.}~\bibnamefont
  {Vasseur}},\ and\ \bibinfo {author} {\bibfnamefont {B.}~\bibnamefont
  {Ware}},\ }\bibfield  {title} {\bibinfo {title} {Stability of superdiffusion
  in nearly integrable spin chains},\ }\href {https://arxiv.org/abs/2102.02219}
  {\bibfield  {journal} {\bibinfo  {journal} {arXiv:2102.02219}\ } (\bibinfo
  {year} {2021})}\BibitemShut {NoStop}%
\bibitem [{\citenamefont {Weiner}\ \emph {et~al.}(2020)\citenamefont {Weiner},
  \citenamefont {Schmitteckert}, \citenamefont {Bera},\ and\ \citenamefont
  {Evers}}]{Weiner2020}%
  \BibitemOpen
  \bibfield  {author} {\bibinfo {author} {\bibfnamefont {F.}~\bibnamefont
  {Weiner}}, \bibinfo {author} {\bibfnamefont {P.}~\bibnamefont
  {Schmitteckert}}, \bibinfo {author} {\bibfnamefont {S.}~\bibnamefont
  {Bera}},\ and\ \bibinfo {author} {\bibfnamefont {F.}~\bibnamefont {Evers}},\
  }\bibfield  {title} {\bibinfo {title} {High-temperature spin dynamics in the
  {{Heisenberg}} chain: {{Magnon}} propagation and emerging
  {{Kardar}}-{{Parisi}}-{{Zhang}} scaling in the zero-magnetization limit},\
  }\href {https://doi.org/10.1103/PhysRevB.101.045115} {\bibfield  {journal}
  {\bibinfo  {journal} {Phys. Rev. B}\ }\textbf {\bibinfo {volume} {101}},\
  \bibinfo {pages} {045115} (\bibinfo {year} {2020})}\BibitemShut {NoStop}%
\bibitem [{\citenamefont {Fava}\ \emph {et~al.}(2020)\citenamefont {Fava},
  \citenamefont {Ware}, \citenamefont {Gopalakrishnan}, \citenamefont
  {Vasseur},\ and\ \citenamefont {Parameswaran}}]{Fava2020}%
  \BibitemOpen
  \bibfield  {author} {\bibinfo {author} {\bibfnamefont {M.}~\bibnamefont
  {Fava}}, \bibinfo {author} {\bibfnamefont {B.}~\bibnamefont {Ware}}, \bibinfo
  {author} {\bibfnamefont {S.}~\bibnamefont {Gopalakrishnan}}, \bibinfo
  {author} {\bibfnamefont {R.}~\bibnamefont {Vasseur}},\ and\ \bibinfo {author}
  {\bibfnamefont {S.~A.}\ \bibnamefont {Parameswaran}},\ }\bibfield  {title}
  {\bibinfo {title} {Spin crossovers and superdiffusion in the one-dimensional
  {{Hubbard}} model},\ }\href {https://doi.org/10.1103/PhysRevB.102.115121}
  {\bibfield  {journal} {\bibinfo  {journal} {Phys. Rev. B}\ }\textbf {\bibinfo
  {volume} {102}},\ \bibinfo {pages} {115121} (\bibinfo {year}
  {2020})}\BibitemShut {NoStop}%
\bibitem [{\citenamefont {Pr{\"a}hofer}\ and\ \citenamefont
  {Spohn}(2000)}]{Prahofer2000}%
  \BibitemOpen
  \bibfield  {author} {\bibinfo {author} {\bibfnamefont {M.}~\bibnamefont
  {Pr{\"a}hofer}}\ and\ \bibinfo {author} {\bibfnamefont {H.}~\bibnamefont
  {Spohn}},\ }\bibfield  {title} {\bibinfo {title} {Universal distributions for
  growth processes in $1+1$ dimensions and random matrices},\ }\href
  {https://doi.org/10.1103/PhysRevLett.84.4882} {\bibfield  {journal} {\bibinfo
   {journal} {Phys. Rev. Lett.}\ }\textbf {\bibinfo {volume} {84}},\ \bibinfo
  {pages} {4882} (\bibinfo {year} {2000})}\BibitemShut {NoStop}%
\bibitem [{\citenamefont {Family}\ and\ \citenamefont
  {Vicsek}(1985)}]{Family1985}%
  \BibitemOpen
  \bibfield  {author} {\bibinfo {author} {\bibfnamefont {F.}~\bibnamefont
  {Family}}\ and\ \bibinfo {author} {\bibfnamefont {T.}~\bibnamefont
  {Vicsek}},\ }\bibfield  {title} {\bibinfo {title} {Scaling of the active zone
  in the {Eden} process on percolation networks and the ballistic deposition
  model},\ }\href {https://doi.org/10.1088/0305-4470/18/2/005} {\bibfield
  {journal} {\bibinfo  {journal} {J. Phys. A: Math. Gen.}\ }\textbf {\bibinfo
  {volume} {18}},\ \bibinfo {pages} {L75} (\bibinfo {year} {1985})}\BibitemShut
  {NoStop}%
\bibitem [{Ion()}]{IonPaper}%
  \BibitemOpen
  \href@noop {} {}\bibinfo {note} {{M. K. Joshi, F. Kranzl, A. Schuckert, I.
  Lovas, C. Maier, R. Blatt, M. Knap, C. F. Roos, Observing emergent
  hydrodynamics in a long-range quantum magnet, to appear in the same arXiv
  posting}}\BibitemShut {NoStop}%
\bibitem [{\citenamefont {Weitenberg}\ \emph {et~al.}(2011)\citenamefont
  {Weitenberg}, \citenamefont {Endres}, \citenamefont {Sherson}, \citenamefont
  {Cheneau}, \citenamefont {Schau{\ss}}, \citenamefont {Fukuhara},
  \citenamefont {Bloch},\ and\ \citenamefont {Kuhr}}]{Weitenberg2011}%
  \BibitemOpen
  \bibfield  {author} {\bibinfo {author} {\bibfnamefont {C.}~\bibnamefont
  {Weitenberg}}, \bibinfo {author} {\bibfnamefont {M.}~\bibnamefont {Endres}},
  \bibinfo {author} {\bibfnamefont {J.~F.}\ \bibnamefont {Sherson}}, \bibinfo
  {author} {\bibfnamefont {M.}~\bibnamefont {Cheneau}}, \bibinfo {author}
  {\bibfnamefont {P.}~\bibnamefont {Schau{\ss}}}, \bibinfo {author}
  {\bibfnamefont {T.}~\bibnamefont {Fukuhara}}, \bibinfo {author}
  {\bibfnamefont {I.}~\bibnamefont {Bloch}},\ and\ \bibinfo {author}
  {\bibfnamefont {S.}~\bibnamefont {Kuhr}},\ }\bibfield  {title} {\bibinfo
  {title} {Single-spin addressing in an atomic {{Mott}} insulator},\ }\href
  {https://doi.org/10.1038/nature09827} {\bibfield  {journal} {\bibinfo
  {journal} {Nature}\ }\textbf {\bibinfo {volume} {471}},\ \bibinfo {pages}
  {319} (\bibinfo {year} {2011})}\BibitemShut {NoStop}%
\bibitem [{\citenamefont {Sherson}\ \emph {et~al.}(2010)\citenamefont
  {Sherson}, \citenamefont {Weitenberg}, \citenamefont {Endres}, \citenamefont
  {Cheneau}, \citenamefont {Bloch},\ and\ \citenamefont {Kuhr}}]{Sherson2010}%
  \BibitemOpen
  \bibfield  {author} {\bibinfo {author} {\bibfnamefont {J.~F.}\ \bibnamefont
  {Sherson}}, \bibinfo {author} {\bibfnamefont {C.}~\bibnamefont {Weitenberg}},
  \bibinfo {author} {\bibfnamefont {M.}~\bibnamefont {Endres}}, \bibinfo
  {author} {\bibfnamefont {M.}~\bibnamefont {Cheneau}}, \bibinfo {author}
  {\bibfnamefont {I.}~\bibnamefont {Bloch}},\ and\ \bibinfo {author}
  {\bibfnamefont {S.}~\bibnamefont {Kuhr}},\ }\bibfield  {title} {\bibinfo
  {title} {Single-atom-resolved fluorescence imaging of an atomic {{Mott}}
  insulator},\ }\href {https://doi.org/10.1038/nature09378} {\bibfield
  {journal} {\bibinfo  {journal} {Nature}\ }\textbf {\bibinfo {volume} {467}},\
  \bibinfo {pages} {68} (\bibinfo {year} {2010})}\BibitemShut {NoStop}%
\bibitem [{\citenamefont {Pertot}\ \emph {et~al.}(2010)\citenamefont {Pertot},
  \citenamefont {Gadway},\ and\ \citenamefont {Schneble}}]{Pertot2010}%
  \BibitemOpen
  \bibfield  {author} {\bibinfo {author} {\bibfnamefont {D.}~\bibnamefont
  {Pertot}}, \bibinfo {author} {\bibfnamefont {B.}~\bibnamefont {Gadway}},\
  and\ \bibinfo {author} {\bibfnamefont {D.}~\bibnamefont {Schneble}},\
  }\bibfield  {title} {\bibinfo {title} {Collinear four-wave mixing of
  two-component matter waves},\ }\href
  {https://doi.org/10.1103/PhysRevLett.104.200402} {\bibfield  {journal}
  {\bibinfo  {journal} {Phys. Rev. Lett.}\ }\textbf {\bibinfo {volume} {104}},\
  \bibinfo {pages} {200402} (\bibinfo {year} {2010})}\BibitemShut {NoStop}%
\bibitem [{\citenamefont {Lühmann}\ \emph {et~al.}(2012)\citenamefont
  {Lühmann}, \citenamefont {Jürgensen},\ and\ \citenamefont
  {Sengstock}}]{Luhmann2012}%
  \BibitemOpen
  \bibfield  {author} {\bibinfo {author} {\bibfnamefont {D.-S.}\ \bibnamefont
  {Lühmann}}, \bibinfo {author} {\bibfnamefont {O.}~\bibnamefont
  {Jürgensen}},\ and\ \bibinfo {author} {\bibfnamefont {K.}~\bibnamefont
  {Sengstock}},\ }\bibfield  {title} {\bibinfo {title} {Multi-orbital and
  density-induced tunneling of bosons in optical lattices},\ }\href
  {https://doi.org/10.1088/1367-2630/14/3/033021} {\bibfield  {journal}
  {\bibinfo  {journal} {New J. Phys.}\ }\textbf {\bibinfo {volume} {14}},\
  \bibinfo {pages} {033021} (\bibinfo {year} {2012})}\BibitemShut {NoStop}%
\bibitem [{\citenamefont {Doyon}(2020)}]{Doyon2020}%
  \BibitemOpen
  \bibfield  {author} {\bibinfo {author} {\bibfnamefont {B.}~\bibnamefont
  {Doyon}},\ }\bibfield  {title} {\bibinfo {title} {Lecture notes on
  generalised hydrodynamics},\ }\href@noop {} {\bibfield  {journal} {\bibinfo
  {journal} {SciPost Phys. Lect. Notes 18}\ } (\bibinfo {year}
  {2020})}\BibitemShut {NoStop}%
\bibitem [{\citenamefont {Ilievski}\ \emph {et~al.}(2018)\citenamefont
  {Ilievski}, \citenamefont {De~Nardis}, \citenamefont {Medenjak},\ and\
  \citenamefont {Prosen}}]{Ilievski2018}%
  \BibitemOpen
  \bibfield  {author} {\bibinfo {author} {\bibfnamefont {E.}~\bibnamefont
  {Ilievski}}, \bibinfo {author} {\bibfnamefont {J.}~\bibnamefont {De~Nardis}},
  \bibinfo {author} {\bibfnamefont {M.}~\bibnamefont {Medenjak}},\ and\
  \bibinfo {author} {\bibfnamefont {T.}~\bibnamefont {Prosen}},\ }\bibfield
  {title} {\bibinfo {title} {Superdiffusion in one-dimensional quantum lattice
  models},\ }\href {https://doi.org/10.1103/PhysRevLett.121.230602} {\bibfield
  {journal} {\bibinfo  {journal} {Phys. Rev. Lett.}\ }\textbf {\bibinfo
  {volume} {121}},\ \bibinfo {pages} {230602} (\bibinfo {year}
  {2018})}\BibitemShut {NoStop}%
\end{thebibliography}%
\clearpage


\setcounter{equation}{0}
\setcounter{figure}{0}
\setcounter{table}{0}
\renewcommand{\theequation}{S\arabic{equation}}
\renewcommand{\thefigure}{S\arabic{figure}}
\renewcommand{\thetable}{S\arabic{table}}

\section*{Supplementary Information}


\section{\label{sec:exp-details}Experimental details}

In this section, we describe our experimental methods (including the sequences for Mott-insulator and initial spin-state preparation). We detail and discuss the data analysis, and show calibration measurements.

\subsection{Mott-insulator preparation}

\begin{figure*}
    \centering
    \includegraphics[scale=1]{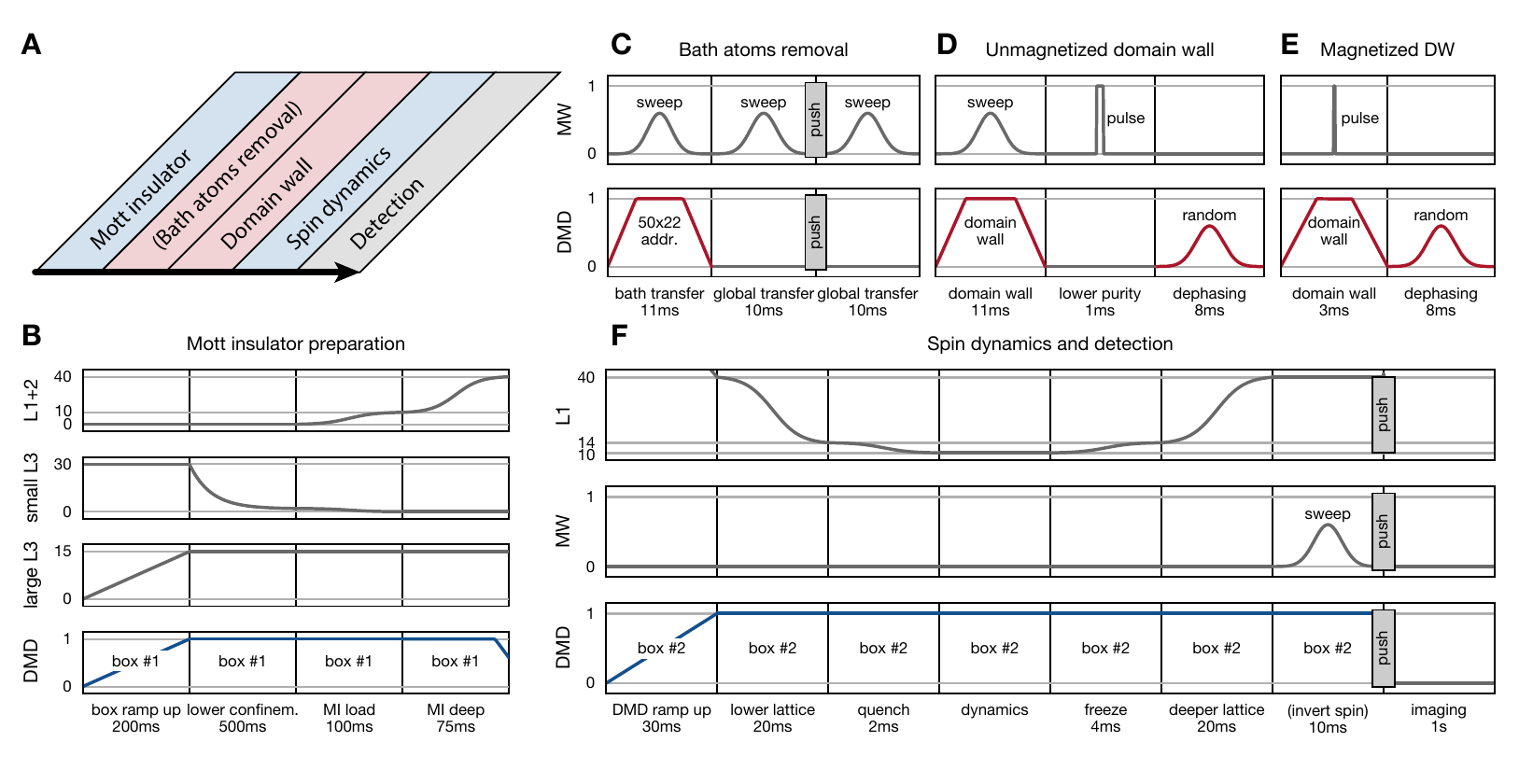}
    \caption{\textbf{Schematic of the experimental sequence.}
        (\textbf{A}) Overview over the stages of a single sequence.
        (\textbf{B}) The Mott insulators are prepared in a box potential. ``L1+2'', ``small L3'' and ``large L3'' denote the horizontal, small vertical and large vertical lattice depths in units of $E_r$. ``DMD'' denotes the DMD potential depth (blue: blue-detuned spin-independent potential), normalized to the power required to form the box potential together with the large L3. ``Box \#1'' denotes projecting the box potential with finite transversal potential barrier.
        (\textbf{C}) The bath atoms are only removed in the 2D measurements. ``MW'' denotes the microwave power (not to scale). The red DMD lines indicate the spin-dependent potential.
        (\textbf{D}, \textbf{E}) Domain-wall preparation sequence for the unmagnetized and magnetized measurements, respectively (see also Fig.~\ref{fig:dw-sequence}).
        (\textbf{F}) For the spin dynamics initialization quench, the lattice in the 1D systems direction (L1) is ramped down to $14\,E_r$, then a $\SI{2}{\milli\second}$ s-shape ramp quenches the system to the target lattice depth (typ. $10\,E_r$). ``Box \#2'' denotes projecting the box potential with high potential barriers. After holding this configuration for the evolution time, the dynamics is frozen with a $\SI{4}{\milli\second}$ s-shape ramp to $14\,E_r$. Depending on the spin state to be detected, an additional global MW transfer flips the spins, followed by a resonant push-out on the spin-down atoms.
    }
    \label{fig:sequence}
\end{figure*}

We started the preparation of our quantum gas for spin dynamics with a two-dimensional (2D) Bose-Einstein condensate of $^{87}\mathrm{Rb}$ atoms in the state $\left|\uparrow\right> = \left|F=1, m_F=-1\right>$, trapped in a single anti-node of a small vertical lattice beam (waist $\SI{70}{\micro\metre}$, lattice constant $a=\SI{532}{\nano\metre}$) (Fig.~\ref{fig:sequence}).
We then ramped up a box potential (denoted ``box \#1'' in Fig.~\ref{fig:sequence}) formed by temporally incoherent light at $\SI{670}{\nano\metre}$ shaped with a digital micromirror device (DMD) and a large vertical lattice beam with lower transverse confinement (waist $\SI{300}{\micro\metre}$), which was kept constant at a depth of $15\,E_r$ throughout the preparation sequence.
Here, $E_r=h^2/8ma^2$ denotes the recoil energy scale characteristic for the lattice.
We avoided spin-dependent potentials by linearly polarizing the light at $\SI{670}{\nano\metre}$.

To fill the box, the depth of the small lattice beam was adiabatically decreased within $\SI{500}{ms}$ to $2\,E_r$.
Afterwards, the small lattice was turned off in $\SI{100}{ms}$ while simultaneously ramping up both horizontal lattice beams to $10\,E_r$, close to the phase transition to a unity-filling Mott insulator.
At this point, the repulsive on-site interaction energy reached $\SI{500}{Hz}$.
The box-potential walls transverse to the lattice along which we probed spin dynamics had a barrier height of $\sim \SI{300}{Hz}$ and a thickness of $7$ sites, over which the potential tapered down to $\sim \SI{100}{Hz}$.
This repulsion sufficed for surplus atoms to leave the system center over the lower transversal potential walls, leaving behind a unity-filling Mott insulator within the box potential.
The atomic distribution was then frozen by ramping up the horizontal lattices to $40\,E_r$, creating arrays of $50\,\times\,22$ sites with a filling of $n_0 = 0.93(1)$.
In the region of interest of $48\,\times\,14$ sites used in the analysis, the average density inhomogeneity was below $\SI{2}{\percent}$.

We manipulated the spins using light at the tune-out wavelength ($\sigma^-$ polarized at $\SI{787.55}{nm}$), controlled spatially with the same DMD used to create the box potential~\cite{Weitenberg2011,Fukuhara2013}.
For local spin flips between $\ket{\uparrow}$ and $\left|\downarrow\right> = \left|F=2, m_F=-2\right>$, we applied differential light shifts of $\SI{40(5)}{kHz}$ and inverted the unaddressed spins by MW transfer.
We detected the density distribution of atoms in the spin-up state with single-atom sensitivity and single-site resolution by resonantly pushing spin-down atoms out before taking a fluorescence image~\cite{Sherson2010}.
Similarly, we detected spin-down atoms by inverting the spins with a microwave (MW) sweep before the push-out.

\subsection{Domain-wall preparation and spin dynamics \label{subsec:dw-dyn}}

\begin{figure*}
    \centering
    \includegraphics[scale=1]{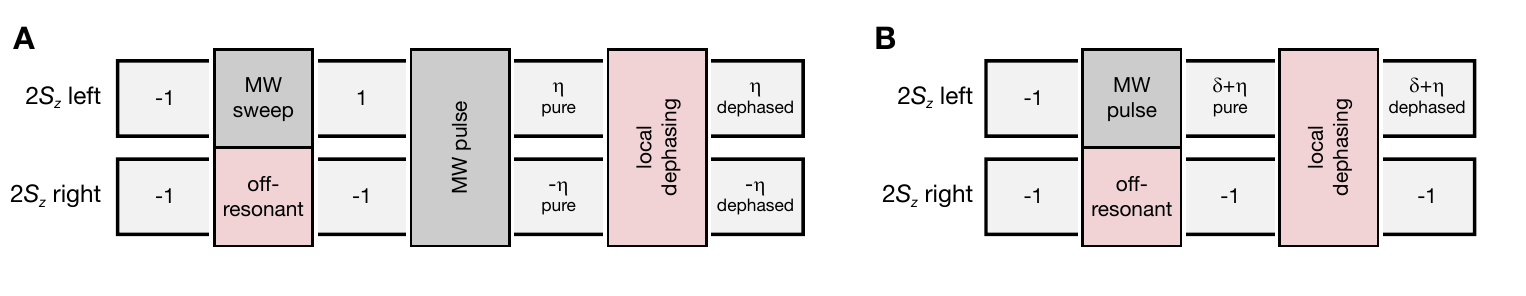}
    \caption{\textbf{Domain-wall preparation sequence.}
        Initial domain-wall preparation starts in the $\left|\downarrow\right>$ polarized state in a deep lattice. The values in the light gray boxes denote the targeted average spin $2 S^z$ on the left and the right side of the domain wall, respectively. Dark-coloured boxes indicate operations performed on the atoms on each domain wall side.
        (\textbf{A}) Unmagnetized domain-wall preparation. A fully polarized domain wall is created using the tune-out DMD laser to shift the spin-flip transition. An adiabatic MW sweep transfers the spins on the unaddressed side. Then a global MW pulse reduces the $S^z$ domain wall contrast and a spatially random tune-out DMD potential locally dephases all spins.
        (\textbf{B}) (Fully) magnetized domain wall preparation. Since one domain wall side is kept polarized, the contrast reducing MW pulse is only applied to one side. As before, the dephasing potential is applied subsequently.
    }
    \label{fig:dw-sequence}
\end{figure*}

The initial spin domain wall was prepared at a lattice depth of $90\,E_r$ in all three directions (Fig.~\ref{fig:dw-sequence}).
For the measurements in 2D, we optically removed all atoms outside of the box potential before beginning the dynamics.
In the measurements without net magnetization, we addressed one half of the system and transferred the remaining atoms.
While the probability for preparing the right spin state exceeded $0.99$ in the bulk of each domain, the probability decreased to about $0.8$ on the sites next to the domain wall.
This is explained by the relative positional drifts of the lattice and the diffraction-limited softening of the DMD-projected pattern.
To prepare domain walls with tunable purities, we subsequently applied a global MW pulse to rotate the spins by a controlled angle.
We then projected a random differential potential varying from shot to shot and from site to site to induce spatially uncorrelated dephasing.
For the measurements with a non-vanishing net magnetization, we use a MW pulse instead of an adiabatic sweep while projecting the differential domain-wall potential to create the state with the desired magnetization. 

We initiated spin dynamics by decreasing the depth of the longitudinal horizontal lattice within $\SI{2}{ms}$ with an s-shape ramp from $14\,E_r$ to $10.0(1)\,E_r$, which was typically used for spin dynamics.
The dynamics took place within a box potential with high walls (denoted ``box \#2'' in Fig.~\ref{fig:sequence}) in both dimensions, at a vertical lattice depth of $15\,E_r$ and in a homogeneous magnetic field;
for the measurements in the 1D cases, the transverse horizontal lattice was kept at $40\,E_r$.
For detection, the dynamics were frozen by increasing the horizontal lattice depth to $14\,E_r$ within $\SI{4}{ms}$.
Subsequently, we measured the parity of the site-resolved occupation via fluorescence imaging~\cite{Sherson2010}.
To ensure that heating was sufficiently small, we performed a reference measurement of the atomic density without spin resolution.
As losses are negligible on the dynamics timescales, a (parity-projected) reduction in density can be identified with the excitation of doublon and holes.
In the 1D case the measured average density drops by up to $\SI{1.5}{\percent}$. In the 2D case it drops by $\SI{4}{\percent}$, where $\SI{2}{\percent}$ can be attributed to atoms tunnelling out of the box potential.

\subsection{Data analysis}

\begin{figure*}
    \centering
    \includegraphics[scale=\figscale]{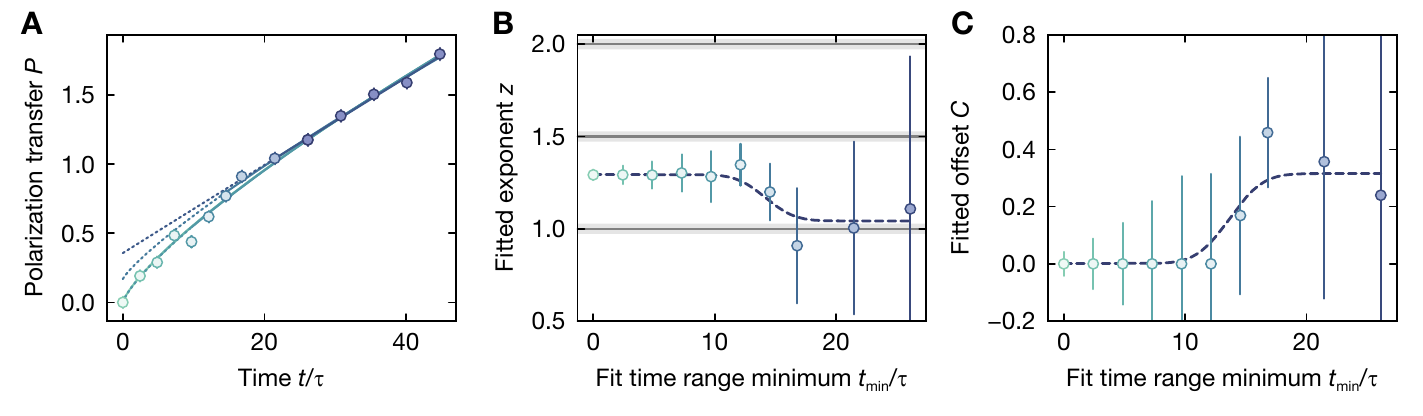}
    \caption{\textbf{Dynamical exponent extraction for the magnetized case.}
        (\textbf{A}) Polarization transfer for magnetized measurement with domain-wall contrast $\eta = 0.12$ and net magnetization $\delta = 0.80$. The lines indicate power-law fits with vertical intercept, $P (t) = A t^{1/z} + C$. The fits are performed over the range $t_\mathrm{min}$ to $t_\mathrm{max} = 45 \tau$, where $t_\mathrm{min}$ is color-coded.
        (\textbf{B}) Fitted exponent $z$ and (\textbf{C}) fitted vertical intercept $C$, where we enforced $C \geq 0$ to prevent unphysical negative $C$. In each case, the dashed lines serve as a guide to the eye, revealing a crossover time located around $t/\tau \sim 14$; for the exponents quoted in the main text, we chose a fitting time range starting at $t_\mathrm{min} > 16 \tau$.
        Error bars denote s.d. of the fit.
    }
    \label{fig:pol-transfer-ballistic}
\end{figure*}

\begin{figure}
    \centering
    \includegraphics[scale=\figscale]{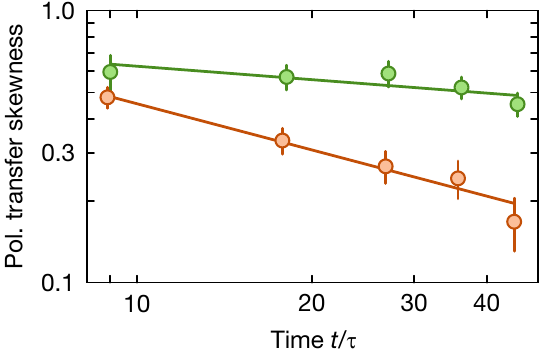}
    \caption{\textbf{Unsubtracted statistics of transferred atoms.}
        Skewness calculated from central moments without initial-state moments subtraction, in comparison to the subtracted skewness shown in Fig.~\ref{fig:5}A.
        The striking difference of the skewnesses between the 1D (green) and 2D case (orange, inter-chain coupling $J_\perp / J = 0.25$) is also visible for unsubtracted central moments. The overall higher skewness results from imperfect initial states comprising nonzero polarization transfer at $t=0$.
        Heuristic power-law fits (lines) yield exponents of $-0.16(9)$ in 1D and $-0.6(1)$ in 2D.
    }
    \label{fig:kpz-distribution-unsubtr}
\end{figure}

From an ensemble of snapshots of the spin distribution, we obtain the average spin densities $n_i^{\uparrow (\downarrow)} (t)$ and calculate the spin profiles as $2 S_i^z (t) = n_i^\uparrow (t) - n_i^\downarrow (t)$.
Both domain-wall contrast $\eta$ and magnetization $\delta$ are extracted by fitting a step function to the $t = 0$ profile.
We obtain the polarization transfer by integrating the deviation of the spin profiles at time $t$ from the initial profile on either side of the domain wall (L, R),
\begin{equation}
    P_\mathrm{L,R} (t) = 2 \sum_{i \in \mathrm{DW}_\mathrm{L,R}} (S_i^z (t) - S_i^z (0)),
    \label{eq:pol-transfer}
\end{equation}
and averaging over the two sides, $P (t) = (P_L (t) - P_R (t)) / 2$ (with the appropriate sign).
To extract the dynamical exponent, we fit the result to a power law, $P (t) \propto t^{1 / z}$.

For the low-purity (1D) measurements (Figs.~\ref{fig:1}, \ref{fig:2}, \ref{fig:4}), we average over about $1000$ 1D shots per point in time; for the high-purity polarization-transfer measurements (Figs.~\ref{fig:1}, \ref{fig:3}) over about $200$ (1D and 2D); for the distribution analysis (Fig.~\ref{fig:5}) over about $3000$ (1D) and $2000$ (2D).

Here we summarize analysis details to the figures in the main text:
\begin{itemize}
    \item Fig.~\ref{fig:1}: For the exponent extraction (Fig.~\ref{fig:1}A inset) in the superdiffusive and diffusive cases we fit a simple power law to the polarization transfer curves; see also the descriptions of Fig.~\ref{fig:2} and \ref{fig:3}, respectively. In the magnetized case, there is a theoretically expected crossover from initially superdiffusive to late-time ballistic transport~\cite{Gopalakrishnan2019}. To account for this and to find a suitable time range for fitting, we fit the power law with vertical intercept and vary the fitted initial time $t_\mathrm{min}$ (Fig.~\ref{fig:pol-transfer-ballistic}). There is a clear qualitative change in behavior at $t_\mathrm{min}/\tau \sim 14$, which suggests a fitting time window starting at $t_\mathrm{min}/\tau > 16$.
    \item Fig.~\ref{fig:2}: For the collapse of the rescaled spin profiles (Fig.~\ref{fig:2}C), we performed 3-site binning to reduce noise.
    \item Fig.~\ref{fig:3}: For the 2D measurements, we observed that the global density drops faster over time than in the 1D case (see sec.~\ref{subsec:dw-dyn}), which could systematically alter the extracted dynamical exponent. To test the impact of this effect, we extract the transport exponents on spin profiles normalized to the average total density, $n_i^\sigma (t) \to n_i^\sigma (t) / \avg{n_i^\uparrow (t) + n_i^\downarrow (t)}_i$. Using this approach, we obtain a modified exponent $z = 2.04(5)$ (instead of $2.08(4)$) for the most affected fully 2D case ($J_\perp = J$), which is on the order of our experimental precision. For 1D data the differences are negligible.
    \item Fig.~\ref{fig:4}: For the spin profiles shown in the color plot (Fig.~\ref{fig:4}A), we show the difference between the time-evolved spin profile and the step-function fit of the initial profile, $S_i^z (t) - S_i^{z,0}$. The fit is used instead of the experimental initial profile $S_i^z (t=0)$ to suppress the impact of noise at $t=0$. Note that this treatment is not necessary for the polarization transfer because the involved integration intrinsically suppresses noise.\\
    When determining the normalized ballistic polarization transfer velocity as a function of net magnetization $\delta$ (Fig.~\ref{fig:4}B), we assume that transport is ballistic after $t_\mathrm{min}/\tau = 16$, see description for Fig.~\ref{fig:1}. We fit a linear function with vertical intercept and use the slope as transport velocity.
    \item Fig.~\ref{fig:5}: When analyzing the polarization transfer statistics, we cannot detect both spin components in a single shot and have to resort to single-species statistics. In each shot (and for each spin species), we instead analyze the transferred atoms $N_T^{\uparrow (\downarrow)}$, i.e. the number of atoms on the side of the domain wall that was initialized with the opposite spin. Assuming that the hole defect fluctuations are small and uniform, this quantity is proportional to the polarization transfer and features the same statistics.\\
    Following this method, we extract the first three central statistical moments $\mu_k$; the skewness as a measure of the asymmetry of the distribution is defined as $\tilde{\mu}_3 = \mu_3 / \mu_2^{3/2}$. The uncertainties are estimated by a bootstrap analysis using the same number for samples and resamples as experimental shots available. We average over the statistics of both spin species. In Fig.~\ref{fig:kpz-distribution-unsubtr}, we plot the bare statistics, where we can clearly distinguish between the strongly skewed 1D measurements and the 2D measurements, whose skewness vanishes with time.\\
    The 1D saturation value, however, exceeds the expected skewness of the Tracy-Widom (TW) distribution, indicating a systematic effect, which we mostly attribute to preparing imperfect initial states. Assuming statistical independence of the polarization transfer due to such initial-state effects and those due to spin dynamics, we correct the central moments at later times by subtracting the moments at $t=0$. We numerically checked that this subtraction is valid in the case of fluctuating domain-wall positions as well as in the case of preparing mixed initial states. The latter case indeed reproduces the approach toward the TW skewness from above.\\
    In Fig.~\ref{fig:5} we therefore show the \emph{subtracted} statistics and analyze the extracted moments through power-law fits. For the 1D case, we confirm that the mean scales superdiffusively with $\alpha = 0.67(1) \approx 1/z = 2/3$, and the standard deviation with $\beta = 0.31(1) \approx 1 / 2z = 1/3$. To distinguish the underlying Heisenberg transport mechanism from linear transport processes, we analyze the skewness and obtain a saturation value of $\tilde{\mu}_3 = 0.33(8) \approx 0.294$, consistent with the value expected for the TW distribution and supporting our conclusion that the transport equations are non-linear. Additionally, we observe that the (subtracted) skewness grows at early times, which resembles numerical simulations, see also Fig.~\ref{fig:MPSMoments}.
\end{itemize}

\subsection{Calibration of the Heisenberg parameters}

As discussed in the main text, the spin-exchange coupling is given by the expression
\begin{align}
    J = \frac{4\tilde{t}^2}{U}.
\end{align}
For our system, the intra- and inter-species scattering lengths give rise to an anisotropy of $\Delta = 0.986$~\cite{Pertot2010}, such that we can consider isotropic coupling, $J_{xy} \approx J_z = J$.


\begin{figure}
    \centering
    \includegraphics[scale=\figscale]{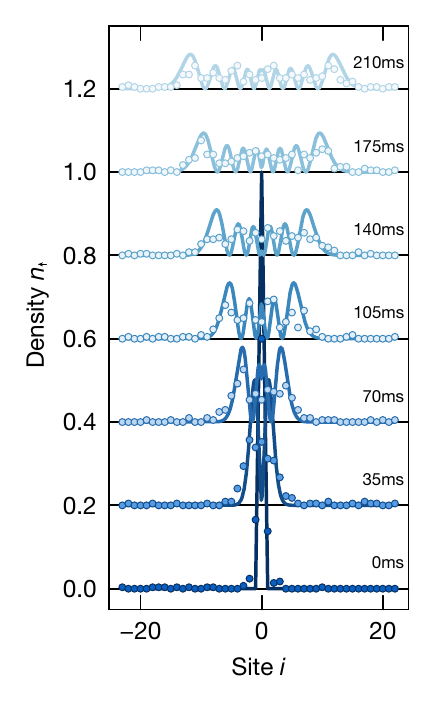}
    \caption{\textbf{Spin quantum walk as a direct measurement of $J$.}
        Single-spin quantum walk at a lattice depth of $10\,E_r$ after different evolution times. The spin-exchange energy $J / h = \SI{10.2 (2)}{Hz}$ is extracted by fitting the analytical expression for the density evolution.
    }
    \label{fig:quantum-walk}
\end{figure}

To obtain the ab-initio values of $\tilde{t}$ and $U$ at a specific lattice depth, we numerically diagonalize the problem of a non-interacting particle in a periodic potential. Considering only the lowest-band contributions, we then obtain the Bose-Hubbard parameters
\begin{align}\begin{split}
    \tilde{t} &= \int w^*(x-a) \left( -\frac{\hbar^2}{2m}\frac{d^2}{dx^2}+V_x \sin^2(k x) \right) w(x)\, dx,\\
    U &= \frac{4\pi\,\hbar^2 a_{\mathrm{s}}}{m}\, \int\,|w(\mathbf{r}) |^4 d^3r.
\end{split}\end{align}
Here $w(\mathbf{r})$ are the Wannier functions for the lowest band.
In order to reach quantitative agreement with our experimental results, we include an additional correction due to bond-charge-induced hopping~\cite{Luhmann2012},
\begin{align}
    \tilde{t}_\mathrm{BC} = \tilde{t} - \frac{4\pi\,\hbar^2 a_{\mathrm{s}}}{m}
    \int w^*(\mathbf{r}-\mathbf{a})\,w^*(\mathbf{r}) w(\mathbf{r}) w(\mathbf{r})\,d^3r.
\end{align}
In Tab.~\ref{tab:BH} we show calculated values of $\tilde{t}$, $U$ and $\tilde{t}_\mathrm{BC}$ for two configurations of lattice depths $(V_1, V_2, V_3)$ relevant for the experiments described in the main text and the Supplementary Information. The lattice depths are calibrated with $1$-$\SI{2}{\percent}$ uncertainty by parametrically heating the gas at $10\,E_r$ for the horizontal lattices and $20\,E_r$ for the vertical lattice.
From the Bose-Hubbard parameters one can calculate the corrected exchange coupling 
\begin{align}
    J_\mathrm{BC} = \frac{4\,\tilde{t}_\mathrm{BC}^2}{U}.
\end{align}

\begin{table}
    \centering
    \setlength{\extrarowheight}{2pt}
    \begin{tabular}{|c|c|c|c|}
        \hline
        $(V_1, V_2, V_3) (E_r)$ & $\tilde{t}/h$ (Hz)& $U/h$ (Hz) & $\tilde{t}_\mathrm{BC}/h$ (Hz) \\
        \hline
        (10, 40, 15)            & 38.91             & 737.0    & 42.70                          \\
        (8, 40, 15)             & 62.50             & 686.3    & 65.73                          \\
        \hline
    \end{tabular}
    \caption{\textbf{Calculated Bose-Hubbard parameters.}
        With lattice depths $V_{1,2,3}$, tunneling energy $\tilde{t}$, interaction energy $U$ and bond-charge-corrected tunneling energy $\tilde{t}_\mathrm{BC}$.
    }
    \label{tab:BH}
\end{table}

\begin{table}
    \centering
    \setlength{\extrarowheight}{2pt}
    \begin{tabular}{|c|c|c|c|}
        \hline
        $(V_1, V_2, V_3) (E_r)$ & $J_\mathrm{0}/h$ (Hz) & $J_\mathrm{BC}/h$ (Hz) & $J_\mathrm{exp}/h$(Hz) \\
        \hline
        (10, 40, 15)            &  8.22                 &  9.90                  & 10.2 (2)               \\
        (8, 40, 15)             & 22.77                 & 27.53                  & 27.0 (2)               \\
        \hline
    \end{tabular}
    \caption{\textbf{Calculated and experimentally measured values of $J$.}
        $J_\mathrm{0}$ and $J_\mathrm{BC}$ denote the calculated exchange coupling without and with the bond-charge term.
    }
    \label{tab:HP}
\end{table}


To check the accuracy of the calculated spin-exchange coupling, we can directly measure $J$ by performing a single-spin quantum walk~\cite{Fukuhara2013} in our system (Fig.~\ref{fig:quantum-walk}). To this end, after preparing the spin-polarized box Mott insulator, we use DMD addressing to flip the central spin in each Heisenberg chain. We then quench the system into the same conditions as used for the domain-wall measurements and measure the single-spin density, which is fitted to the expected time-evolving density $n_i^\uparrow (t) = \mathcal{J}_i^2 (J t / \hbar)$, where $\mathcal{J}_i$ denotes the Bessel function of the first kind. The measurements yield $J_\mathrm{exp} / h = \SI{10.2 (2)}{Hz}$ and $\SI{27.0 (2)}{Hz}$ at lattice depths of $10\,E_r$ and $8\,E_r$, respectively, and are compared to the calculated values in Tab.~\ref{tab:HP}.

\subsection{Spin state transfers}

The preparation of each sequence requires up to six spin transfers, which are realized by adiabatic MW sweeps. We characterize the transfer fidelity by subsequently performing tens of transfers and pushing the majority spin component out, yielding a transfer probability of $0.9996 (1)$. The initial spin state purity in combination with the push fidelity is $0.999 (1)$.

As we are measuring average atom number transfers across the domain wall with a precision of $\sim 0.1$ atoms, involuntary transfers due to off-resonant scattering of the projected $\SI{787}{nm}$ addressing light have to be suppressed. We measure the number of spin transfers of $\ket{\uparrow}$ atoms after illuminating the atoms for up to $\SI{500}{ms}$ to estimate the scattering rate. For the $\SI{8}{ms}$ addressing time, we obtain a probability of less than $\SI{20}{\percent}$ for a scattering event in the 1D chain. Addressing is always performed in a deep lattice of $90\,E_r$ in all directions to prevent heating in the motional degrees of freedom.
Furthermore, we prevent scattering-induced spin flips by addressing $\ket{\downarrow} = \ket{F=2, m_F=-2}$ atoms, which scatter on the cycling transition and do not decay into other hyperfine states..


\section{\label{sec:add-meas}Additional measurements}

In this section, we show additional measurements, where we cross-check our experiments with spin-spiral initial states, analyze effects of Hubbard dynamics, and verify that magnetic gradients did not have significant influence on our results.

\subsection{Transport measurement with spin spiral}

\begin{figure}
    \centering
    \includegraphics[scale=\figscale]{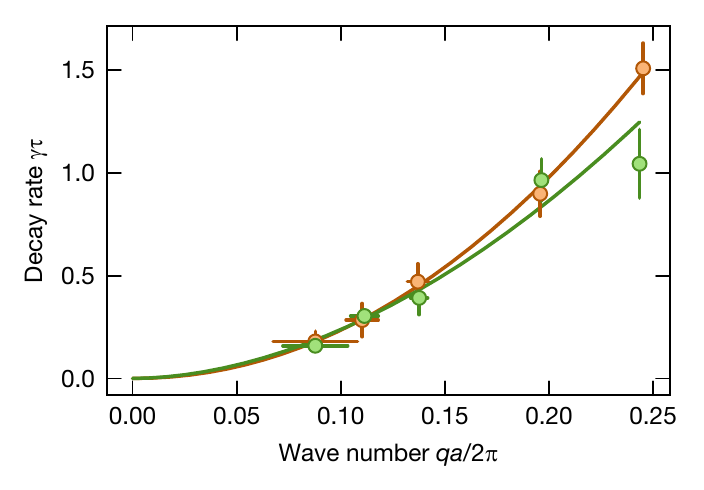}
    \caption{\textbf{Power law from spin spiral decay.}
        Decay rate $\gamma$ of the $g_{n^\uparrow}^{(2)}$ correlator visibility of spin spirals with wave number $q$ at a lattice depth of $10\,E_r$ (green) and $8\,E_r$ (orange). Both measurements give an exponent consistent with diffusive transport of $z = 1.9 (2)$ and $2.0 (1)$, respectively.
        Error bars denote s.d. of the fits.
    }
    \label{fig:spin-spiral}
\end{figure}

Spin transport in Heisenberg chains has been studied previously in ultracold atomic systems by measuring the contrast decay of spin spiral states ~\cite{Hild2014,Jepsen2020}, where in both cases a dynamical exponent consistent with diffusion ($z \approx 2$) was extracted for the isotropic Heisenberg point.

For a direct comparison with the domain-wall initial state, we studied the decay dynamics of longitudinal spin spirals. Analogous to the previous experiment in this setup~\cite{Hild2014}, we prepared the spiral state via a Ramsey sequence, where a linear magnetic gradient imprinted a spiral pattern during the time between the pulses. We obtained the decay rates $\gamma$ by fitting an exponential function to the visibility, $V (t) \propto e^{-\gamma t}$, of the second-order correlation function $\avg{\hat{n}_i^\uparrow (t)\,\hat{n}_{i+d}^\uparrow (t)} $. Performing this measurement for varying spin spiral wave numbers $q$ allows us to fit the transport power law $\gamma \propto q^z$.

\begin{figure*}
    \centering
    \includegraphics[scale=\figscale]{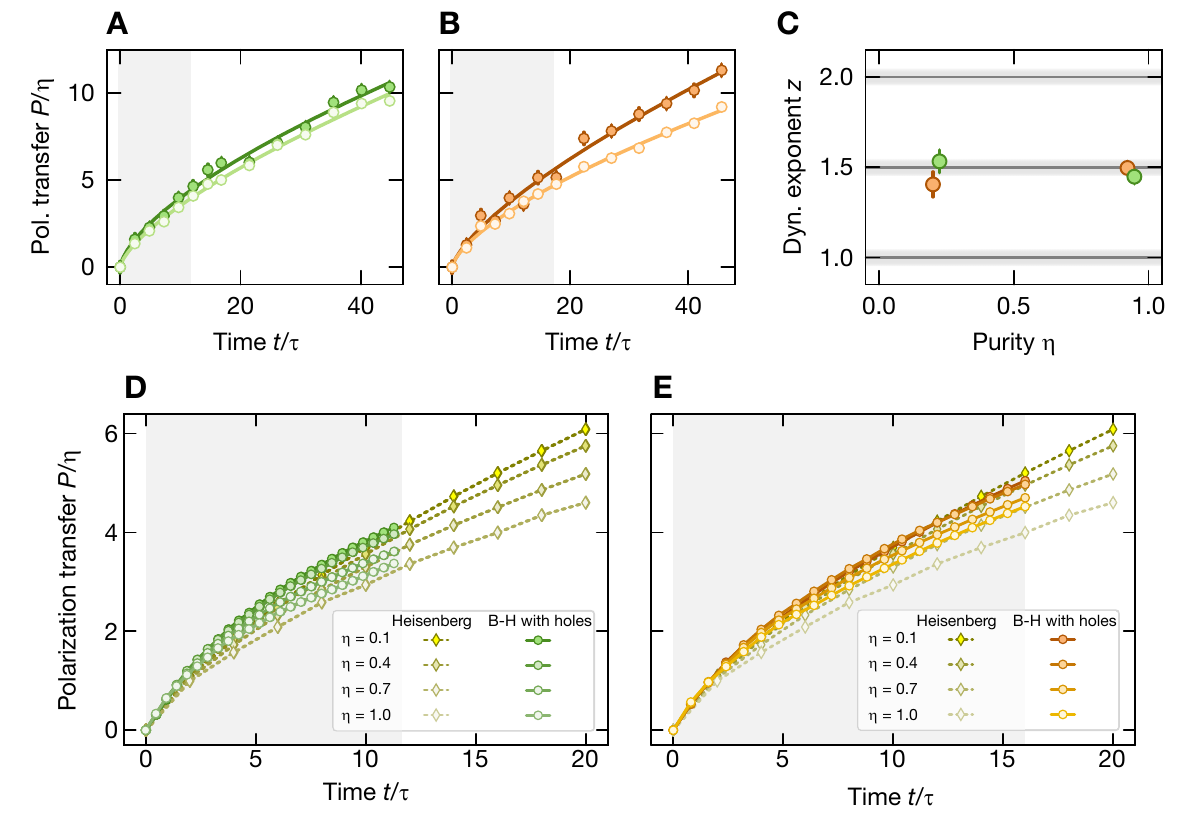}
    \caption{\textbf{Polarization transfer for varying domain-wall purity and lattice depth.}
        Experimental measurements at a lattice depth of $10\,E_r$ (\textbf{A}, $U/\tilde{t} \sim 17$) and $8\,E_r$ (\textbf{B}, $U/\tilde{t} \sim 10$) for domain-wall purities $\eta \sim 0.2$ (dark) and $\eta \sim 0.9$ (light) in the 1D unmagnetized case. The extracted exponent (\textbf{C}) does not show a significant dependence on purity; the normalized polarization transfer is similarly insensitive to $\eta$. Error bars denote s.d. of the fit.
        Numerical simulations of the two-species Bose-Hubbard model with $\SI{7}{\percent}$ holes for a lattice depth of $10\,E_r$ (\textbf{D}) and $8\,E_r$ (\textbf{E}), in comparison to ideal Heisenberg model simulations. The shaded regions mark the time window, where numerical Bose-Hubbard simulations were feasible.
        While the Heisenberg model features smaller normalized polarization transfer for higher purities, the Bose-Hubbard numerics confirm experimental observations that Bose-Hubbard effects lead to weaker purity dependence. However, the simulations suggest stronger dependence on $\eta$ when the lattice depth increases, which does not completely agree with the experimental observation and remains a question for future work.
    }
    \label{fig:purity-ldepth}
\end{figure*}

We can indeed reproduce the diffusive behavior observed in prior work~\cite{Hild2014} for both lattice depths of $10\,E_r$ and $8\,E_r$ (Fig.~\ref{fig:spin-spiral}), which deviates significantly from the superdiffusive exponents extracted from the direct polarization transport measurements. 
This highlights that the dynamics of high-purity spin spiral states can be very distinct from the underlying (high-temperature) universal transport exhibited by the Heisenberg model. 
A theoretical analysis of these special states, and the universality of their dynamics, remain an intriguing open question.

\subsection{Bose-Hubbard model effects}

Our experiment realizes the two-species Bose-Hubbard model
\begin{align} \begin{split}
    \hat{H} = &-\tilde{t} \sum_{\avg{i,j}, \sigma} \hat{c}_{i,\sigma}^\dagger \hat{c}_{j,\sigma}^{\mathstrut} 
     + \frac{U}{2} \sum_{i,\sigma} \hat{n}_{i,\sigma} \left(\hat{n}_{i,\sigma} - 1\right)\\
     & +U \sum_i \ \hat{n}_{i, \uparrow}\,\hat{n}_{i, \downarrow},
\end{split} \end{align}
with tunneling energy $\tilde{t}$ and on-site interaction energy $U$. At unit filling, $\avg{\hat{n}_i} = 1$, the Heisenberg model emerges perturbatively in $\mathcal{O} (\tilde{t} / U)$ in the deep lattice limit with spin-exchange energy $J = 4 \tilde{t} (\tilde{t} / U)$. The Heisenberg spin operators are then given by $\hat{S}_j^x + i \hat{S}_j^y = \hat{a}_{\uparrow,j}^\dagger \hat{a}_{\downarrow,j}$, $\hat{S}_j^x - i \hat{S}_j^y = \hat{a}_{\downarrow,j}^\dagger \hat{a}_{\uparrow,j}$ and $\hat{S}_j^z = (\hat{n}_{\uparrow,j} - \hat{n}_{\downarrow,j}) / 2$, where $\hat{a}_{\sigma,j}^{(\dagger)}$ denote ladder and $\hat{n}_{\sigma,j}$ number operators for spin $\sigma$ on site $j$.

In the experiment, thermal defects in the form of holes and doublons are important. Due to the lower energy cost, hole defects are predominant and the average filling is generally less than unity. We measure typical parity-projected filling fractions of $n_0 = 0.93 (1)$, where $1$-$\SI{2}{\percent}$ of the defects can be explained by imaging artefacts. Further defects may be introduced when quenching the lattice depth, as this generates doublon-hole fluctuations in the charge sector.


In Fig.~\ref{fig:purity-ldepth}A-C we compare 1D unmagnetized polarization transfer measurements at different purities and lattice depths. The low-purity curves agree well with each other and Heisenberg numerics. Whereas Heisenberg simulations indicate a substantial reduction of the normalized polarization transfer for initial domain walls with increasing purity $\eta$ (see also sec.~\ref{subsec:num-purity}), experimental measurements show a significantly weaker dependence on purity.
The smaller dependence at $10\,E_r$ indicates that $\tilde{t}$-timescale effects might be responsible for the transport corrections.
The reduced dependence is also visible in numerical two-species Bose-Hubbard simulations (Figs.~\ref{fig:purity-ldepth}D-E).

Surprisingly, all power law exponents extracted from data or numerics do not show a strong dependence on purity and are close to $z = 3/2$, the value theoretically expected in the infinite-temperature limit. For the ideal pure case at $\eta=1$, polarization transfer is expected to become eventually diffusive with logarithmic corrections~\cite{Gamayun2019}; however, in the presence of experimental imperfections and at the timescales accessible in our experiment, superdiffusion appears to be remarkably robust.

\subsection{Magnetic gradient effects}

\begin{figure}
    \centering
    \includegraphics[scale=\figscale]{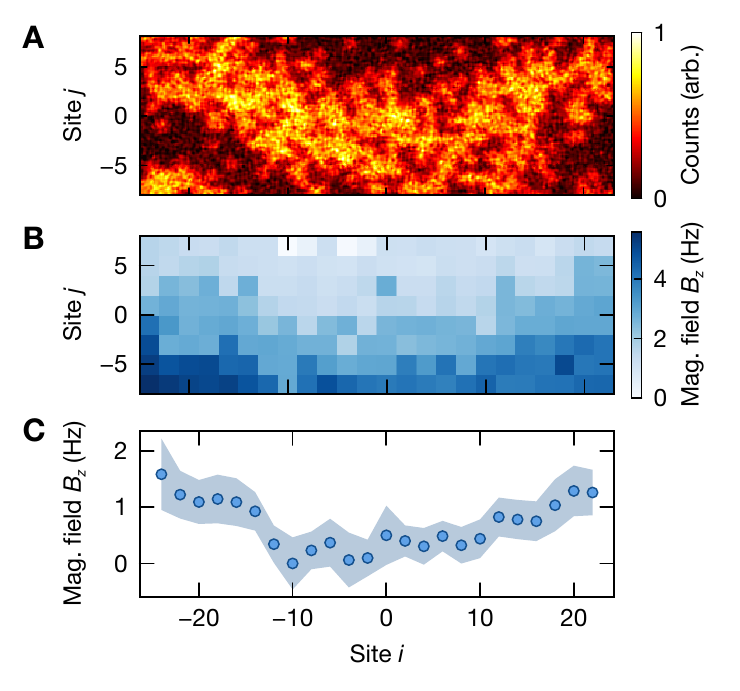}
    \caption{\textbf{Residual magnetic gradients.}
        The residual magnetic gradients are extracted by fitting the interference fringes in a Ramsey experiment.
        (\textbf{A}) Single-shot example of a $\SI{300}{\milli\second}$ dark time Ramsey measurement.
        (\textbf{B}) Magnetic field difference map in the analysis region obtained by fitting local fringes to a series of Ramsey measurements with dark times up to $\SI{400}{\milli\second}$. The field was optimized to minimize gradients along the 1D chains.
        (\textbf{C}) Magnetic field distribution along the 1D systems. Maximal local gradients remain below $\SI{0.1}{\hertz}/a \sim 0.01 J/a$ (at $10\,E_r$). The shaded area denotes the standard deviation when averaging over the parallel 1D chains.
    }
    \label{fig:magnetic-field}
\end{figure}

\begin{figure*}
    \centering
    \includegraphics[scale=\figscale]{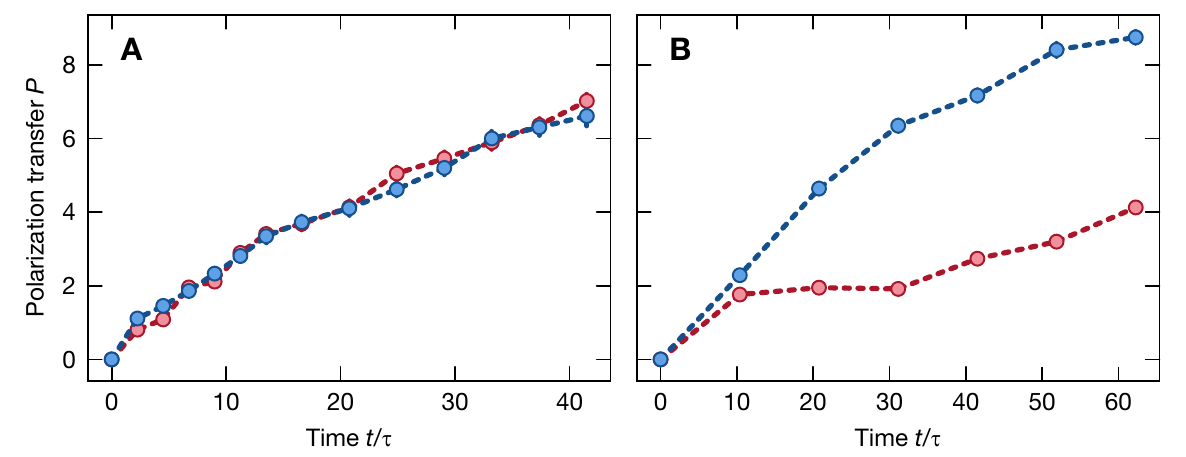}
    \caption{\textbf{Polarization transfer with magnetic gradient.}
        Polarization transfer of pure domain walls at a lattice depth of $8\,E_r$. The two colors indicate mutually inverted domain walls (i.e. $\left|\downarrow\right>$ atoms are on the left or on the right side).
        (\textbf{A}) For minimized magnetic gradient, both spin configuration display identical behavior.
        (\textbf{B}) At a magnetic gradient of $\SI{1.4}{\hertz}/a \sim 0.05 J/a$, a strong difference is visible. Due to the gradient, one initial state constitutes a relatively low-energy state, while the other state has high energy. The low-energy state remains largely localized and has a suppressed polarization transfer.
    }
    \label{fig:grad-pol-transfer}
\end{figure*}

As we work with large system sizes and long evolution times, we have to minimize spatial inhomogeneities of the effective magnetic field $B_i^z$ in the Heisenberg model, $\hat{H}' = \hat{H} + 2 \sum_i B_i^z \hat{S}_i^z$. The effective field comprises the differential light shift due to the trapping light and the actual magnetic field as the two hyperfine spin states employed in our work feature a differential Zeeman energy shift of $\SI{2.1}{kHz/mG}$. Hence we used linearly polarized $\SI{670}{\nano\metre}$ light for the box potential and moved the in-plane magnetic field minimum close to the atoms.

We mapped out the effective magnetic field by performing a Ramsey sequence formed by two MW $\pi/2$-pulses, interspersed with a hold time of up to $\SI{400}{\milli\second}$ (Fig.~\ref{fig:magnetic-field}). While global phase coherence is lost within about $\SI{100}{\micro\second}$, relative phases are preserved, and give rise to an interference pattern with globally random phase. The local gradients are typically around $\SI{0.04}{\hertz}/a$ along the 1D chains and small compared to the $10\,E_r$ spin-exchange energies around $\SI{10}{\hertz}$.
We verified that the effective magnetic field remains constant throughout the evolution time.

In order to detect the experimental signatures of such a gradient, we deliberately apply a uniform $\SI{1.4}{\hertz}/a$ gradient along the chains and analyze the polarization transfer. We compare the behavior with a spin-inverted domain wall, which is equivalent to flipping the gradient polarity. We observe that in one configuration, the transfer is sped up while it is slowed down in the opposite configuration (Fig.~\ref{fig:grad-pol-transfer}). Furthermore, the curves develop a strong deviation from a power law. For our transport measurements, we verified that the magnetic field inhomogeneities are negligible and the transport is identical for inverted and non-inverted domain walls, similar to Fig.~\ref{fig:grad-pol-transfer}A.


\section{\label{sec:numerics}Numerical simulations}

\subsection{Methods and convergence}
Throughout the entire work, the numerical simulations presented are performed using density matrix truncation (DMT)~\cite{White2018,Ye2020}, a novel and powerful method which allows us to directly calculate the dynamics up to late times of the mixed state describing the system. 
The method is closely related to the well-known time-evolving block decimation (TEBD) for the time evolution of a wavefunction represented as a matrix product state. 
However, in DMT, the system is described with a \emph{density matrix} using a matrix product density operator (MPDO). 
The time-evolution operator is, similarly, Trotterized into a set of two-site gates which act on the density matrix in MPDO form to simulate a small discrete time step. 
After each evolution step, the MPDO is truncated to a specified maximum bond dimension of the matrix product representation. 
The key feature of DMT is that this truncation is chosen so that it preserves local observables---such as the energy density, magnetization and their currents---which are crucial to capturing the transport of interest. 
Preserving these local observables, rather than maximizing the mutual information (as conventional TEBD does) allows DMT to correctly capture late-time equilibration and hydrodynamics~\cite{Ye2020}.

\begin{figure*}[t!]
    \centering
    \includegraphics[scale=\figscale]{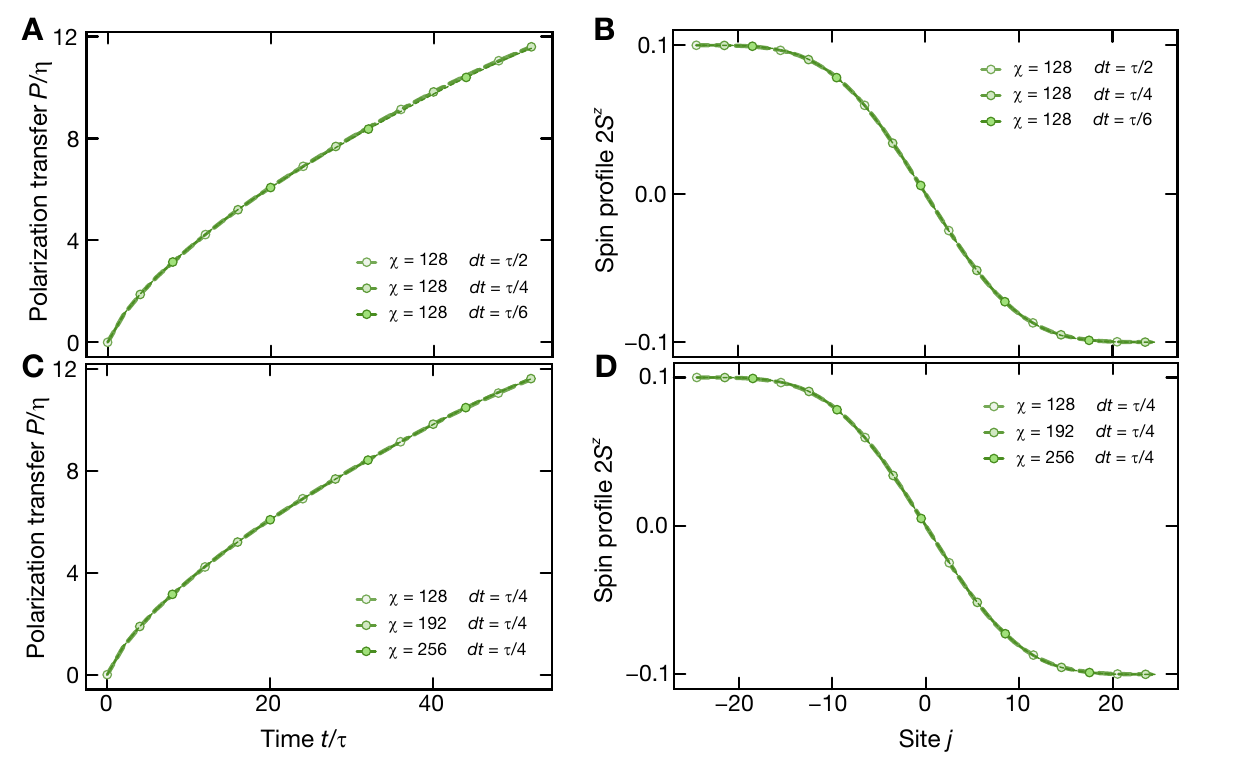}
    \caption{\textbf{Convergence of DMT simulation}.
        (\textbf{A}, \textbf{B}) Convergence with respect to the Trotter step length $dt$.
        (\textbf{C}, \textbf{D}) Convergence with respect to the bond dimension $\chi$.
        The spin profiles are measured at time $t/\tau = 20$. 
        While in this plot we only show the numerics for $\eta=0.1$, we also observe the same convergence in simulation for other parameter values. 
    }
    \label{fig:num-convergence}
\end{figure*}

As we have discussed, in DMT the density matrix is approximated by an MPDO with a maximum bond dimension $\chi$. The Trotterization of the time-evolution operator introduces a further approximation controlled by the Trotter step size $dt$. It is therefore important to verify that our simulations have converged with the values of these meta-parameters used. We do this by simulating the same dynamics with $\chi\in \{128,192,256\}$, and $dt \in \{\tau/2, \tau/4, \tau/6\}$. For both polarization density and total polarization transferred across the domain wall, we observe fast convergence with both bond dimensions and time step, Fig.~\ref{fig:num-convergence}, up to at least time $t/\tau=10^2$.

\subsection{Different transport regimes}

\begin{figure}[t!]
    \centering
    \includegraphics[scale=\figscale]{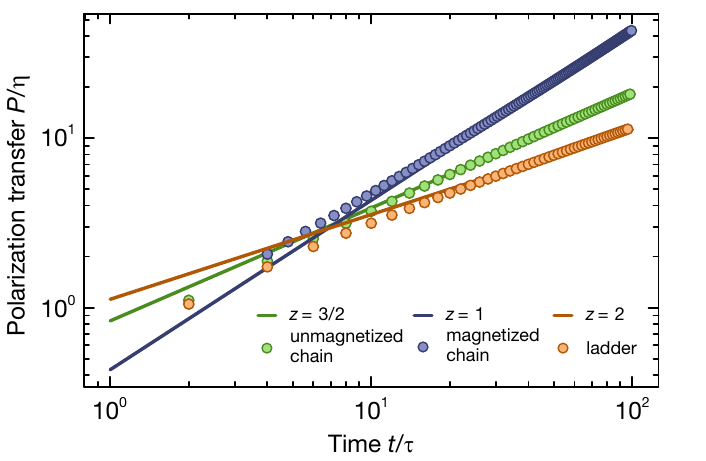}
    \caption{\textbf{Numerical simulation of different setups, highlighting different types of transport.}
        Circles: total polarization transport obtained from the numerical simulation. Lines: fits with different power-law exponents corresponding to different universality classes.
        We choose large enough system sizes ($L=300$ for spin chains and $L=160$ for spin ladders) to avoid finite-size effects within a timescale of $100\tau$. 
    }
    \label{fig:num-regimes}
\end{figure}

Having confirmed that our simulations have converged, we demonstrate that DMT can accurately capture the different transport regimes (diffusive, ballistic and superdiffusive) observed in Heisenberg chains, (Fig.~\ref{fig:num-regimes}). As in the experiment, at zero net magnetization $\delta$, transport is superdiffusive, while at finite $\delta$ it is ballistic. Although the full 2D model in the experiment is not accessible numerically, we can observe diffusion by considering a ``ladder'' system of two coupled Heisenberg chains. Integrability is also broken in this setup, leading to diffusive transport.

\subsection{\label{subsec:num-purity}Effect of purity $\eta$ in the superdiffusive transport}

So far, the numerical simulations were conducted close to infinite temperature in the linear response regime (i.e. at small purity $\eta$) where analytical results are best understood. 
Indeed, the required SU(2) symmetry for the KPZ superdiffusion is only strictly present when $\delta=0$ and $\eta=0$. 
Nevertheless, for finite $\eta$ of the initial state, as the domain wall melts, the magnetization approaches zero and the SU(2) symmetry is restored in the middle of the chain. 
More importantly, the superdiffusive polarization transport is bottlenecked by such unmagnetized region, since finite net magnetization will lead to ballistic transport, which is faster than superdiffusion. Therefore, even for large (but non-unity) purity, it is natural to expect the superdiffusion behavior to still persist. 

\begin{figure}[t!]
    \centering
    \includegraphics[scale=\figscale]{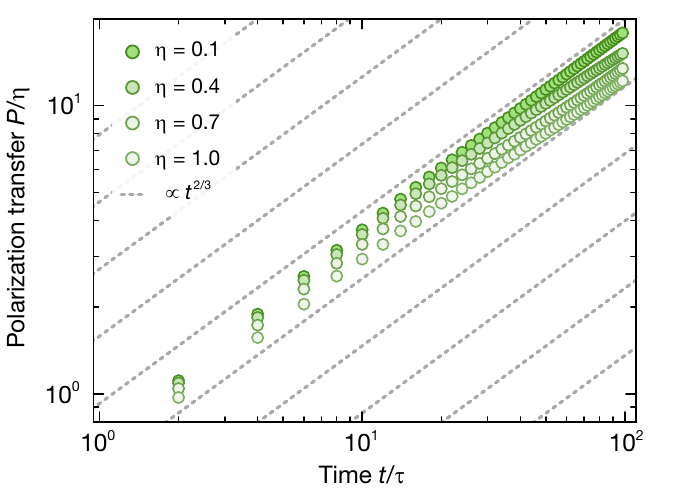}
    \caption{\textbf{Total polarization transfer for initial domain walls with different purity.}
        As the purity $\eta$ increases, the dynamical exponent $z=3/2$ as the diagnostic of superdiffusion persists, while the superdiffusion constant decreases. 
        Gray dashes are guidelines for power-law growth with $z=3/2$. 
    }
    \label{fig:num-vary-eta}
\end{figure}

In Fig.~\ref{fig:num-vary-eta}, we numerically study the effect of purity $\eta$ in the measured polarization transfer. We see that, while the overall magnitude of polarization transferred varies with $\eta$, the associated dynamical exponent $z$ remains consistent with KPZ superdiffusion ($z=3/2$) up until the pure initial state, $\eta < 1$. This is consistent both with theoretical expectation and with the experimental observations of superdiffusion at finite and large $\eta$. 

Curiously, precisely at $\eta = 1$, the behavior is known to be diffusive with logarithmic corrections. However, distinguishing this behavior from superdiffusion, requires following the dynamical evolution to very late time in large system sizes \cite{Ljubotina2017,Misguich2017,Ye2020}.

\subsection{Velocity of ballistic transport}

\begin{figure*}[t!]
    \centering
    \includegraphics[scale=\figscale]{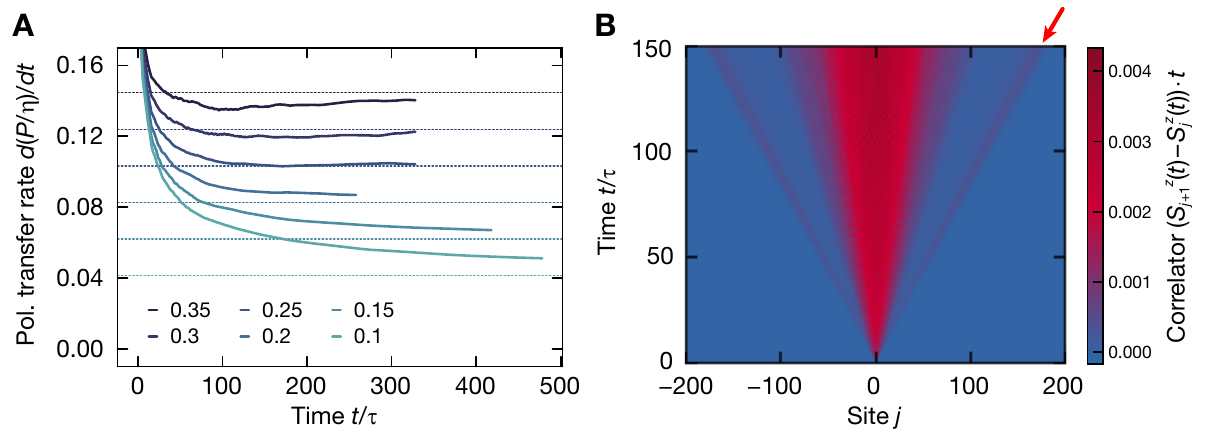}
    \caption{\textbf{Numerical simulation of polarization transport from magnetized initial state.}
        (\textbf{A})  Magnetization transfer as a function of time for different net magnetizations $\delta$. At long times, the saturation value of polarization transfer rates (solid lines) is expected to exhibit a linear dependence on the net magnetization (solid lines).
        (\textbf{B}) By calculating the spatial gradient of the polarization profile, we obtain the spatio-temporal profile of the dynamical structure factor, i.e. the spin-spin correlation function. The inner light cone represents the dominant contribution to the linear transport of the magnetization, while the outer light cone (marked by the red arrow) highlights the linear transport of magnons.
        The magnons only carry net polarization and appear in the dynamical structure factor when the initial state is magnetized.
    }
    \label{fig:num-vary-mag}
\end{figure*}

The velocity of the ballistic polarization transfer at finite net magnetization is theoretically expected to be proportional to the magnetization $\delta$~\cite{Gopalakrishnan2019a}. We numerically test this claim and arrive at good agreement with the analytical expectation (Fig.~\ref{fig:num-vary-mag}A), in agreement with the trend observed in the experiment in Fig.~\ref{fig:4}B. We emphasize that this velocity is not the same as the outer ballistic ``light-cone'' velocity of the fastest-travelling quasiparticles (having the smallest size), which we expect to be independent of $\delta$, but rather corresponds to the velocity associated with the bulk of the spin transport inside the light-cone, as is illustrated in in Fig.~\ref{fig:num-vary-mag}B and further discussed below.

\begin{figure}[t!]
    \centering
    \includegraphics[scale=\figscale]{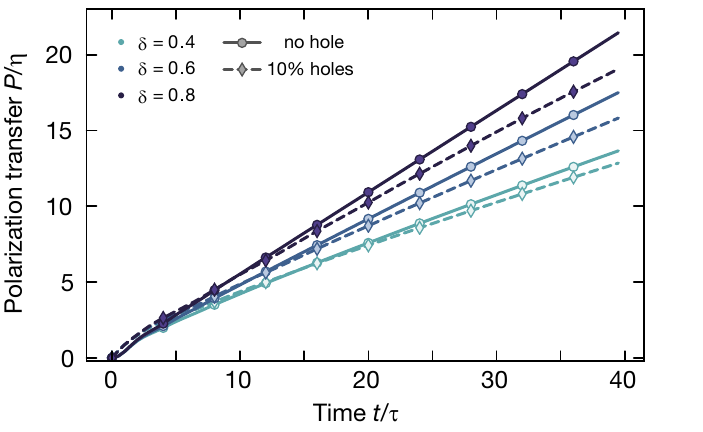}
    \caption{\textbf{Numerical simulation of ballistic transport from magnetized initial states with the presence of holes.}
        In the simulation, we only consider the lowest three energy levels on each site.
        We introduced $10\%$ holes in the initial state and compare the subsequent polarization transfer with the ideal Heisenberg model (which is equivalent to the case without holes).
        Across different values of magnetization, the introduction of holes always slows down the ballistic polarization transport, and delays the crossover time from the superdiffusive regime to the ballistic. 
    }
    \label{fig:ballistic_tJ}
\end{figure}

However, the polarization transfer in experiments is slower than numerical simulation, Fig.~\ref{fig:4}A. The most obvious culprit for this disagreement is the non-vanishing population of holes in the experiment.
To account for the effect of these holes in the simplest possible manner, we simulate the dynamics of the $t$-$J$ model:
\begin{equation}
\begin{split}
    \hat{H}_{t-J}=-\tilde{t}\sum_{i,\sigma}&(\hat{a}^\dag_{i,\sigma} \hat{a}_{i+1,\sigma}+h.c.)\\
    &+J\sum_i\left(\hat{\vec{S}}_i \cdot \hat{\vec{S}}_{i+1}-\frac{\hat{n}_i \hat{n}_{i+1}}{4}\right),
\end{split}
    \label{eq:t-J_Hamiltonian}
\end{equation}
where $\sigma$ is the spin polarization, and $\hat{n}_i=\sum_\sigma a^\dag_{i,\sigma}a_{i,\sigma}$. 
Introducing a small amount of holes into the magnetized ($\delta\neq 0$) initial domain-wall state, we still observe late-time ballistic transport (Fig.~\ref{fig:ballistic_tJ}). 
However, the associated velocity becomes smaller than in the ideal Heisenberg model. 
Furthermore, the crossover time to the ballistic regime is delayed. 
This suggests that the presence of holes in the experiment indeed drives the observed quantitative discrepancy with the numerics for the Heisenberg model, but does not destroy the expected late-time ballistic behavior.

\section{Fluctuations in KPZ dynamics}

One of the key features that distinguishes KPZ dynamics from other $z=3/2$ dynamical processes (such as rescaled diffusion and L\'evy flights) is the non-linearity of the underlying dynamical process.
This non-linearity has an important consequence: fluctuations of the polarization transfer are not symmetric around the mean.

In the case of KPZ dynamics in the 1D Heisenberg model, the magnetization profile $S^z (x, t)$ is mapped to the spatial derivative of the height field $h (x, t)$ of the KPZ equation, $S^z (x, t) \sim \partial_x h (x,t)$~\cite{Ljubotina2019}.
The initial state studied in our work, i.e. the domain wall in magnetization $S^z (x, 0) \sim 2 \Theta (x) - 1$ (with Heaviside function $\Theta$), then maps to a wedge initial state of the height field $h (x, 0) \sim -|x|$.
The polarization transfer $P(t)$, being the spatially integrated magnetization profile $P (t) \sim \int_{-\infty}^0 S^z (x, t)\,dx$, thus maps to the height field at the peak of the wedge $h(0, t)$.
The dynamical fluctuations of precisely this quantity, $h(0, t)$, were numerically studied in Ref.~\cite{Hartmann2018} for a classical lattice model known to be in the KPZ universality class. At late times, these fluctuations showed an approach to the GUE Tracy-Widom (TW) distribution, which (owing to universality) also describes the distribution of largest eigenvalue of random matrices from the Gaussian unitary ensemble (GUE).

This feature provides a path to directly observing the underlying KPZ dynamics. 
Leveraging the access to single experimental snapshots in quantum-gas microscopes, as well as the single-site resolution, we can immediately build the distribution of the polarization transfer and measure the aforementioned asymmetry via the skewness of the distribution.

\subsection{Polarization transfer fluctuations near pure state}

\begin{figure*}
    \centering
    \includegraphics[scale=\figscale]{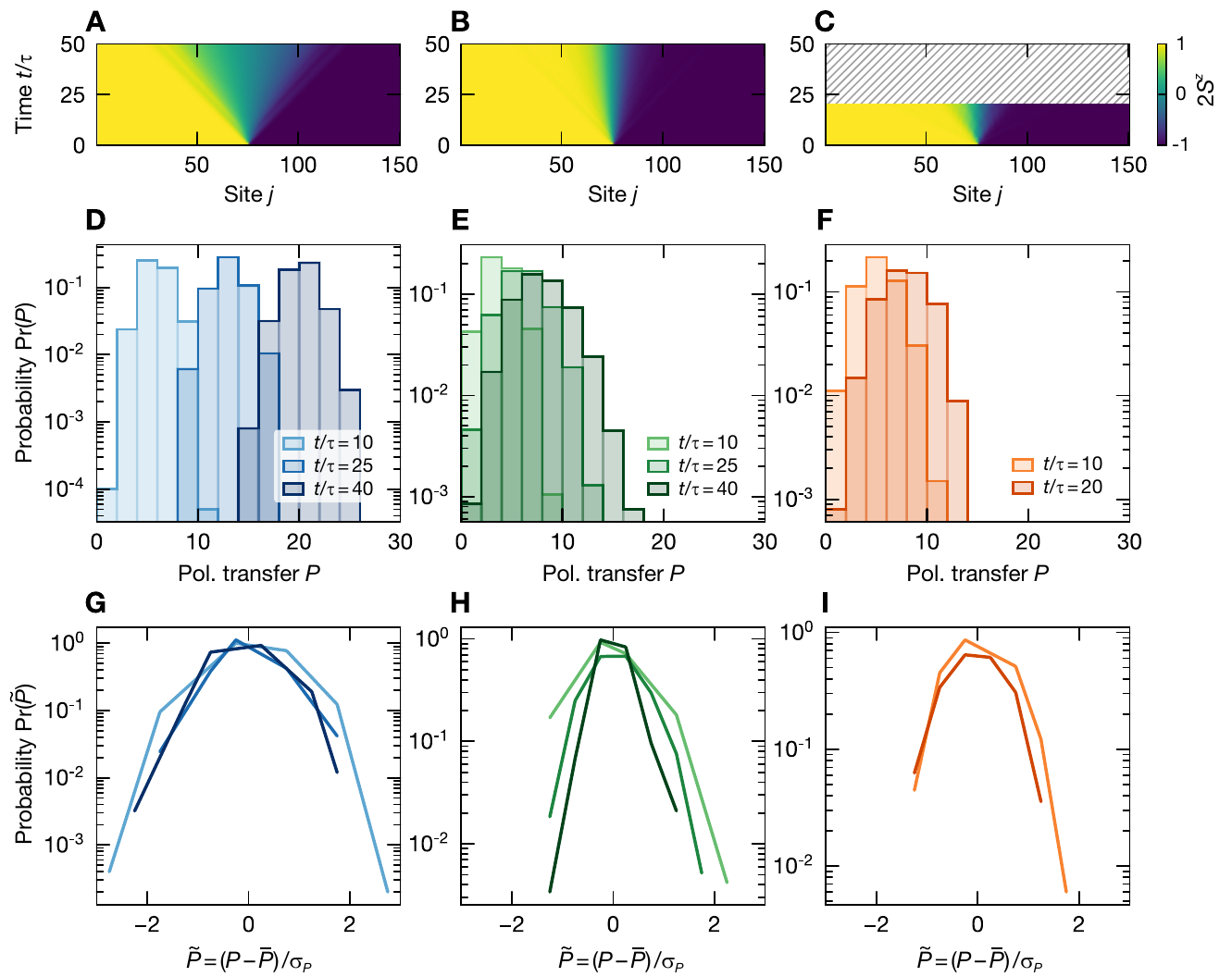}
    \caption{\textbf{Magnetization dynamics in different transport regimes.}
        Ballistic (\textbf{A},\textbf{D},\textbf{G}), KPZ superdiffusive (\textbf{B},\textbf{E},\textbf{H}), diffusive (\textbf{C},\textbf{F},\textbf{I}, simulated up to $t/\tau = 20$) transport for different models initialized in a pure domain wall ($\eta = 1$).
        (\textbf{A}-\textbf{C}) Polarization-profile dynamics for the entire chain as a function of time.
        (\textbf{D}-\textbf{F}) Distribution of the polarization transfer $P$ with respect to the initial state measured by projecting the quantum state into the measurement basis according to the Born rule (akin to the single-shot measurement procedure performed in the experiments).
        (\textbf{G}-\textbf{I}) Rescaled probability distribution $\tilde{P}$ according to the average $\overline{P}$ and standard deviation $\sigma_P$ of the transferred polarization.
    }
    \label{fig:MPSFluctuations}
\end{figure*}

\begin{figure*}
    \centering
    \includegraphics[scale=\figscale]{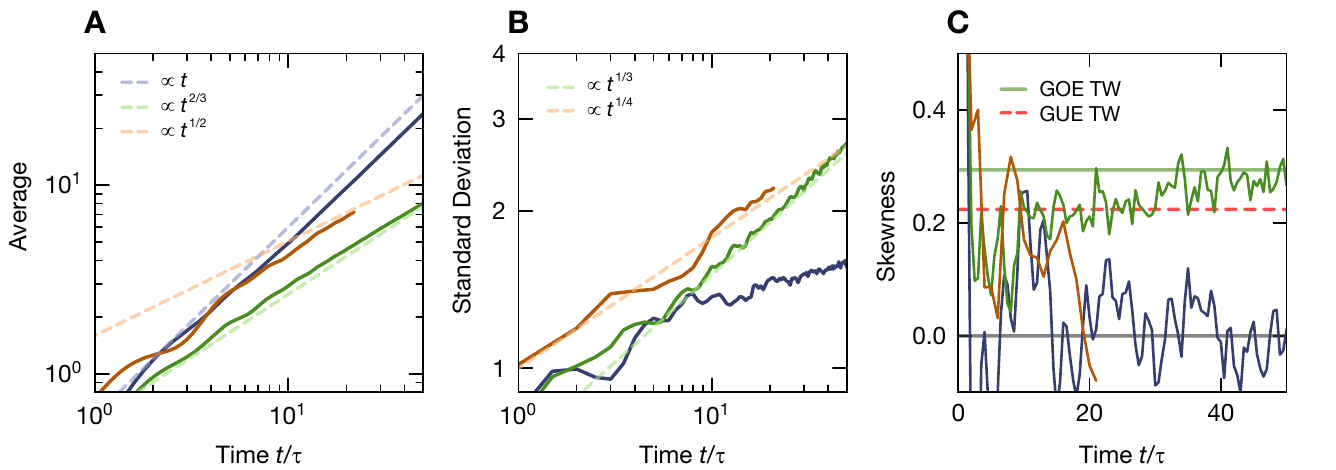}
    \caption{\textbf{Statistical moments of polarization-transfer distribution in different transport regimes.}
        Average (\textbf{A}), standard deviation (\textbf{B}) and skewness (\textbf{C}) of the distribution of transferred polarization $P$ for the different models exhibiting ballistic (blue), KPZ superdiffusive (green), diffusive (orange) transport. Dashed lines indicate expected scaling behavior. In \textbf{C}, the constant lines mark the late-time expectation for KPZ (red dashed for our initial state, green for different one), and for both diffusive and ballistic (black).
    }
    \label{fig:MPSMoments}
\end{figure*}

In order to better highlight the fluctuations of magnetization, we begin by considering the pure domain wall dynamics.
While at the Heisenberg point, the late-time dynamics approaches a logarithmically corrected diffusion \cite{Gamayun2019}, the approach to this universal behavior is very slow and the system exhibits superdiffusion with KPZ characteristics for intermediate timescales.

We then study the dynamics of the pure domain wall under three different Hamiltonians: the Heisenberg Hamiltonian (Eq.~\ref{eq:XXZ}), the easy-plane XXZ model, and a next-nearest neighbor interacting Heisenberg model with Hamiltonian:
\begin{align}\begin{split}
    \hat{H}_{\mathrm{NNN}}/J = \sum_{i} \hat{S}^x_i\hat{S}^x_{i+1} &+ \hat{S}^y_i\hat{S}^y_{i+1} + 1.05 \hat{S}^z_i\hat{S}^z_{i+1} \\
    - 0.764 \sum_{i} \hat{S}^x_i\hat{S}^x_{i+2} &+ \hat{S}^y_i\hat{S}^y_{i+2} + 1.05 \hat{S}^z_i\hat{S}^z_{i+2}
\end{split}\end{align}
These models allow us to display the three universality classes of the dynamics: KPZ, ballistic and diffusive dynamics, respectively.
Using an MPS representation of the quantum state, we perform time evolution of a chain of $150$ spins; using a TEBD algorithm with step size $\delta t = 0.1$ and large enough bond dimension to observe convergence.
Leveraging the MPS representation, average quantities can be directly computed via the expectation value of the corresponding observables (Figs.~\ref{fig:MPSFluctuations}A-C).
At the same time, the single-shot experimental measurement can be simulated by sampling the quantum state over the measurement basis (according to the Born rule).
By computing the number of spins that crossed the initially pure domain wall, we can directly obtain, not only the average polarization profile, but its entire distribution (Figs.~\ref{fig:MPSFluctuations}D-F).

Crucially, the resulting distributions in these three cases look very different. 
In both ballistic and diffusive cases (Figs.~\ref{fig:MPSFluctuations}D,F), the distributions remain symmetric, while in the superdiffusive case (Figs.~\ref{fig:MPSFluctuations}E), the distribution develops a tail towards large polarization transfer.
Such behavior is easier to observe upon subtracting the average $P$ and rescaling with the standard deviation (Figs.~\ref{fig:MPSFluctuations}G-I).

These differences can be quantified by looking at the evolution of the different moments of the distribution: average, standard deviation and (normalized) skewness Fig.~\ref{fig:MPSMoments}.
We note that the average and standard deviation scale differently with time, highlighting the different dynamical exponents.
More importantly, we observe that the skewness for both ballistic and diffusive regimes decays to zero (signifying that the distribution is symmetric), while, in the superdiffusive case, the skewness remains non-zero and approaches a finite value.
While this value approaches the skewness of a Tracy-Widom distribution (green line, Fig.~\ref{fig:MPSMoments}C), it corresponds to the Gaussian-orthogonal-ensemble (GOE) TW distribution (with skewness $\sim 0.293$), which is expected for a different set of initial conditions.
Namely, for the wedge initial configuration we expect the fluctuation distribution to approach the GUE Tracy-Widom distribution whose skewness is $\sim 0.225$ (red dashed line, Fig.~\ref{fig:MPSMoments}C)~\cite{Prahofer2000}.
At present, the origin of this deviation remains unclear.
These numerics highlight two important facts: first, that the skewness can identify the underlying nature of the transport dynamics, and second, that this distinction occurs within the experimentally accessible time.

When $\eta<1$, using DMT, we observe an analogous behavior, although accurately capturing the fluctuations (corresponding to higher-order non-local observables) requires much more extensive numerical resources.


\section{Decay of skewness in linear transport}

In this section we describe why, in linear transport, the skewness of the polarization transfer distribution always decays.
Leveraging the linearity of the transport equations, we can compute the magnetization distribution $F(x_0,t)$ via a convolution of the initial domain, with the Green's function of equation $f(x,t)$ (i.e. the magnetization profile dynamics starting from a delta-function of magnetization at time $t=0$):
\begin{equation}
    F(x_0,t) \propto \int_{-\infty}^0 dx~f(x-x_0,t) = \int_{-\infty}^{x_0} dx~f(x,t).
\end{equation}

If the linear transport has dynamical exponent $z$, $f(x,t)$ (for large enough $x$ and $t$) is given by a scaling function:
\begin{equation}
    f(x,t) = t^{-1/z} g\left( \frac{x}{t^{1/z}}\right),
\end{equation}
which implies that $F(x_0,t) = G(x_0/t^{1/z})$. We note that this exactly corresponds to the rescaling performed in the main text.

This means that all moments of the distribution, which are integrals of powers of $F(x_0,t)$, at fixed time, will scale with $t^{1/z}$. Since the skewness is given by the ratio of the third moment and the second moment to the $3/2$ power, we have that the skewness will decay as $t^{-1/2z}$ and thus become zero at late enough times.
This holds for all higher moments, ensuring that the rescaled distribution of the polarization transfers approaches the normal distribution.


\section{Polarization transfer in the GHD framework}

In what follows we briefly explain how to
compute polarization transfer $P(t)$ within the Generalized Hydrodynamics (GHD) framework, in the limit of weak quenches, $\eta \ll 1$. In this limit, one can express $P(t)$ in terms of linear-response correlation functions. As we will see, this quantity is related to (but subtly different from) the transport coefficients that have previously been calculated in the literature. 

In the $\eta \ll 1$ limit, it is known~\cite{Ljubotina2017} that the magnetization profile $\rho(x,t)=\langle \hat{S}_x^z(t)\rangle$ is related to the linear-response dynamical correlation function $C(x,t) \equiv \langle \hat{S}^z_x(t) \hat{S}^z_0(0) \rangle$ via the expression
\begin{equation}
C(x,t) = \partial_x \rho(x,t)
\end{equation}
where we used the continuum notation for derivatives, for simplicity, although in a lattice model they should strictly be expressed in terms of discrete differences. We can invert this relation to read:
\begin{equation}
\rho(x,t) = \rho(-\infty) + \int_{-\infty}^x dx' C(x',t).
\end{equation}
The polarization transfer is (up to time-independent constants) given by
\begin{equation}\label{pdef}
P(t) = \int_{-\infty}^0 \int_{-\infty}^x dx dx' C(x',t).
\end{equation}
In the hydrodynamic limit, $C(x,t)$ will take the scaling form $t^{-1/z} C(x^z/t)$, where $z$ is the dynamical exponent. By dimensional analysis of the expression for $P(t)$ one can see that this in general implies $P(t) \sim t^{1/z}$. 

We now discuss the scaling of this quantity within GHD in the Heisenberg model, working at $\delta \neq 0$. In the Heisenberg model, there are infinitely many quasiparticle species, labeled by the ``string index'' $s$. Each quasiparticle species propagates ballistically ($z = 1$); the velocity of a quasiparticle (and the density of such quasiparticles $\rho_s(\theta)$) depends on both $s$ and the quasimomentum $\theta$. In terms of these, $C(x,t)$ can be written as~\cite{Doyon2020}
\begin{equation}\label{ghdcorr}
C(x,t) = \sum_s \int d\theta \rho_s(\theta) [m_s^{\mathrm{dr}}(\theta)]^2 \delta(x - v_s^{\mathrm{eff}} (\theta) t).
\end{equation}
This expression can be interpreted as follows: each quasiparticle propagates ballistically with some velocity $v^{\mathrm{eff}}_s(\theta)$ that depends on the nature of the background state, and carries some effective spin (which, again, depends on the background state via dynamical screening). Correlations between the spacetime points $(0,0)$ and $(x,t)$ come from all quasiparticles whose trajectories pass through both spacetime points. The thermodynamic Bethe ansatz provides a framework within which all the quantities appearing in Eq.~\eqref{ghdcorr} can straightforwardly be computed. In the high temperature limit, closed-form expressions exist~\cite{Ilievski2018} for all the quasiparticle data in Eq.~\eqref{ghdcorr}. After some coarse-graining the correlator can be written as~\cite{Gopalakrishnan2019}
\begin{equation}
C(x,t) = \sum_s \frac{1}{\tilde v_s t} \rho_s (m_s^{\mathrm{dr}})^2 \Theta(x - \tilde v_s t)
\end{equation}
where $\tilde v_s$ is some characteristic velocity for quasiparticles of species $s$, and $\Theta$ is the Heaviside step function. Plugging this form into Eq.~\eqref{pdef} we find that
\begin{equation}\label{polsum}
P(t) = t \sum_s \rho_s (m_s^{\mathrm{dr}})^2 \tilde v_s.
\end{equation}
In the Heisenberg model at nonzero $\delta$, one has the following scaling forms. For $s \delta \lesssim 1$, we have $\rho_s \sim 1/s^3$ and $m_s^{dr} \sim \delta s^2$, while for $s \delta \gtrsim 1$, we have $\rho_s \sim \exp(-\delta s)$ and $m_s^{dr} = s$. For all $s$ we have the scaling $v_s \sim 1/s$. Thus the sum over species gets cut off at $s \sim 1/\delta$, yielding the expression
\begin{equation}
P(t) \sim \delta^2 \sum_{s < 1/\delta} O(1) \sim \delta t.
\end{equation}
Superdiffusion can be recovered within this framework by noting that even when $\delta = 0$, fluctuations of $\delta$ cause quasiparticles to move in a time-dependent apparent magnetic field.

It is interesting to contrast the expression~\eqref{polsum} with that for the Drude weight $\mathcal{D}$ (i.e., singular part of the zero-frequency conductivity) of the Heisenberg model:
\begin{equation}\label{drudeweight}
\mathcal{D} = \sum_s \rho_s (m_s^{\mathrm{dr}})^2 |\tilde v_s|^2.
\end{equation}
Because of the extra factor of velocity in Eq.~\eqref{drudeweight} relative to Eq.~\eqref{polsum}, the contribution of slow quasiparticles to the polarization transfer is much larger than their contribution to the Drude weight.

\end{document}